\documentclass[pra,twocolumn,showpacs,preprintnumbers,amsmath,amssymb]{revtex4}

\usepackage{mathrsfs}%hanamoji
\usepackage{graphicx}% Include figure files
\usepackage{dcolumn}% Align table columns on decimal point
\usepackage{bm}% bold math

\begin{document}

%%%%%
%\begin{minipage}{16cm}

\title{Accuracy matrix in generalized simultaneous measurement of a qubit system}
\author{Takahiro Sagawa$^1$}
\author{Masahito Ueda$^{1,2}$}
\affiliation{$^1$Department of Physics, Tokyo Institute of Technology,
2-12-1 Ookayama, Meguro-ku, Tokyo 152-8551, Japan \\
$^2$ERATO Macroscopic Quantum Control Project, JST, 2-11-16 Yayoi, Bunkyo-ku, Tokyo 113-8656, Japan
}
\date{\today}

\begin{abstract}
We formulate the accuracy of a quantum measurement for a qubit (spin-1/2) system in terms of a 3 by 3 matrix.  This matrix, which we refer to as the accuracy matrix, can be calculated  from a positive operator-valued measure (POVM) corresponding to the quantum measurement.  Based on the accuracy matrix, we derive trade-off relations between the measurement accuracy of two or three noncommuting observables of a qubit system.    These trade-off relations offer a quantitative information-theoretic representation of Bohr's principle of complementarity.   They can be interpreted as the uncertainty relations between measurement errors in simultaneous measurements and also as the trade-off relations between the measurement error and  back-action of the measurement.  A no-cloning inequality is derived from the trade-off relations.   Furthermore,  our formulation and the results obtained  can be applied to analyze quantum-state tomography.  We also show that the accuracy matrix is closely related to the maximum-likelihood estimation and the Fisher information matrix for a finite number of samples; the accuracy matrix tells us how accurately we can estimate the probability distributions of observables of an unknown state by a finite number of quantum measurements. 
\end{abstract}


%\keywords{ keywords}%Use showkeys class option if keyword
\maketitle
%%%%%%
%\end{minipage}\vspace{2mm}

\section{Introduction}
Accessible information about a quantum system  is restricted by the noncommutability of observables.  The nature of this restriction  can be classified essentially into two categories:  fluctuations inherent in a quantum system and the  error caused by the process of measurement.  These aspects of uncertainty constitute the two distinctive features  of quantum mechanics.

The Kennard-Robertson uncertainty relation such as $\Delta x \Delta p \geq \hbar /2$ describes  quantum fluctuations that are independent of the measurement process~\cite{Kennard,Robertson,Deutsch,Maassen-Uffink}.  According to Bell's theorem~\cite{Bell}, this type of quantum fluctuations prohibits  us from  presupposing any  ``element of reality''~\cite{Einstein}  behind the probability distributions of observables.  The measurement error, on the other hand, is  determined by the process of  measurement which is characterized  by a positive operator-valued measure (POVM)~\cite{Davies-Lewis,Nielsen-Chuang}. In the idealized  error-free limit,  quantum measurement is  described by projection operators which, however, cannot always be implemented experimentally.

Information about more than one observable can be obtained from  a single POVM in   simultaneous measurement  of two noncommuting observables and quantum-state tomography. It is known, however, that, in simultaneous measurements, at least one of the observables  cannot be measured without incurring a measurement error~\cite{Neumann}. In this context, various uncertainty relations between the measurement errors of noncommuting observables have been studied~\cite{Bohr,Arthurs-Kelly,Busch,Yamamoto-Haus,Arthurs-Goodman,Martens-Muynck,Appleby,Muynck,Busch-Shilladay,Hall,Ozawa-1,Andersson-Barnett-Aspect,Massar,Kurotani,Jammer}. 

In this paper, we quantify the measurement accuracy and the measurement error of observables in terms of a given POVM $\textbf{E} = \{ \hat E_k \}$  by introducing  $3 \times 3$ accuracy matrix $\chi(\textbf{E})$  calculated from the POVM.  Based on this accuracy matrix, we derive trade-off relations between the measurement accuracy of two or three observables, these  being  stronger  trade-off relations than those derived in our previous work~\cite{Kurotani}.  They   can be interpreted as the uncertainty relations between the measurement errors  of noncommuting observables in simultaneous measurements  or as the uncertainty relations between the measurement error and   back-action of the measurement~\cite{Heisenberg,Fuchs,Banaszek-Devetak,Ozawa-2,Busch-Heinomen-Lahti}. In addition, a  no-cloning inequality~\cite{Wootters-Zurek,Dieks,Barnum,Cerf} is derived from the trade-off relations.

In a rather different context, the maximum-likelihood estimation~\cite{Fisher,Lehmann} has been investigated as the  standard scheme of   quantum state tomography for a finite number of samples. Several studies have  focused  on the efficiency and optimality of the  estimation of  an unknown quantum state~\cite{Banaszek,Hradil,Rehacek-2,Thew,James}.  We show that our characterization of the measurement accuracy can be related to the maximum-likelihood estimation and that the accuracy matrix can be interpreted as an average of the Fisher information matrix over the state to be measured.  The trade-off relations can also be interpreted as those concerning  the accuracy of the estimate of various probability distributions of noncommuting observables.

The constitution of this paper is as follows.  In Sec.~I\hspace{-.1em}I, we formulate the general quantum measurement of a qubit (spin-1/2) system.  In Sec.~I\hspace{-.1em}I\hspace{-.1em}I, we define the accuracy matrix and investigate its properties.  Based on this accuracy matrix, we define  the accuracy parameter and error parameter in a particular  direction of measurement.   In Sec.~I\hspace{-.1em}V, we derive the trade-off relations between the accuracy parameters or the error parameters in two or three directions.  In Sec.~V, we apply  the trade-off relations to specific problems: the uncertainty relations between measurement errors in nonideal joint measurements, the uncertainty relations between the error and  back-action,  a no-cloning inequality,  and  quantum state tomography.  In Sec.~V\hspace{-.1em}I, we   point out a close connection between the accuracy matrix and the Fisher information matrix.  We  conclude this paper in Sec.~V\hspace{-.1em}I\hspace{-.1em}I.

\section{Quantum Measurement of a Qubit System}

 We consider a quantum measurement described by POVM $\textbf{E} = \{ \hat E_k \}$ ($k=1,2, \cdots, m$) on  state $\hat \rho$ of a qubit system, where $k$ denotes the outcome of the measurement.  POVM $\textbf{E}$ satisfies $\sum_k \hat E_k = \hat I$, with $\hat I$ being the identity operator, and can be parameterized as 
\begin{equation}
\hat E_k = r_k (\hat I +\bm v_k \cdot \hat{\bm \sigma}),
\end{equation}
where $\hat{\bm \sigma} \equiv (\hat \sigma_x, \hat \sigma_y, \hat \sigma_z)$ represents the Pauli matrices. The requirements that the sum of $\hat E_k$'s equals the identity operator and that all of them be nonnegative are met if and only if
\begin{equation}
\sum_k r_k =1, \ \sum_k r_k \bm v_k = \bm 0, \ r_k > 0 , \ |\bm v_k| \leq 1 \ {\rm for \ all} \ k.
\label{POVM}
\end{equation}
 We can also parameterize density operator $\hat \rho$ as
\begin{equation}
\hat \rho = \frac{1}{2}( \hat I + \bm{s}_0 \cdot \hat{\bm \sigma}),
\end{equation}
where $\bm s_0$ is the Bloch vector satisfying $| \bm s_0 | \leq 1$. Conversely, for a given $\hat \rho$, $\bm s_0$ is calculated as $\bm s_0 = {\rm tr}(\hat \rho \hat{\bm \sigma})$.  The probability of obtaining the measurement outcome $k$ is then given by 
\begin{equation}
q_k \equiv {\rm tr} (\hat E_k \hat \rho) = r_k (1+ \bm v_k \cdot \bm s_0).
\label{qk}
\end{equation}

Any observable $\hat O$ of the qubit system can be diagonalized as
\begin{equation}
\hat O = \lambda_+ \hat P(+;\bm n) + \lambda_- \hat P(-; \bm n),
\end{equation}
where $\lambda_+$ and $\lambda_-$ are the corresponding eigenvalues,  $\hat  P(+;\bm n)$ and $\hat P(+;\bm n)$ are projection operators with $\bm n$ being a three-dimensional unit vector, and
\begin{equation}
\hat P(\pm;\bm n) = \frac{1}{2} (\hat I \pm \bm n \cdot \hat{\bm \sigma}).
\end{equation}
The probability distribution of  observable $\hat O$ is then given by
\begin{equation}
p(\pm; \bm n) \equiv {\rm tr}(\hat P(\pm;\bm n) \hat \rho) = \frac{1}{2} (1 \pm \bm n \cdot \bm s_0).
\label{distribution}
\end{equation}

If we are not interested in  eigenvalues of the observables but are only concerned with the directions ($\pm$) of the outcome, we can replace $\hat O$ with $\bm n \cdot \hat{\bm \sigma}$ by setting $\lambda_{\pm} = \pm 1$.  In the following analysis, we identify  observable $\lambda_+ \hat P(+;\bm n) + \lambda_- \hat P(-; \bm n)$ with the observable $\bm n \cdot \hat{\bm \sigma}$  and  refer to the probability distribution in Eq.~(\ref{distribution}) as that in the direction of $\bm n$.  
 
We discuss three typical  examples.

\textbf{Example 1}
(\textit{projection measurement}).
We can precisely measure $\bm n \cdot \hat{\bm \sigma}$ by the projection measurement described by the POVM $\textbf{E} = \{ \hat P( +;\bm n), \hat P(-;\bm n) \}$.

\textbf{Example 2} 
(\textit{Nonideal measurement}).
A more general class of measurements can be described by the POVM $\textbf{E}$ consisting of two positive operators parametrized as
\begin{equation}
\hat E(+;\bm n) = r (\hat I + \varepsilon_1 \bm n \cdot \hat{\bm \sigma}), \ \hat E(-;\bm n) = (1-r) (\hat I - \varepsilon_2 \bm n \cdot \hat{\bm \sigma}),
\label{nonideal1}
\end{equation}
where $\bm n$ is a unit vector, $r\varepsilon_1 -(1-r)\varepsilon_2 =0$, $0<r<1$, $-1 \leq \varepsilon_1 \leq 1$, and $-1 \leq \varepsilon_2  \leq 1$.  This POVM corresponds to a nonideal measurement of the observable $\bm n \cdot \hat{\bm \sigma}$~\cite{Martens-Muynck,Muynck}. It can be reduced to a projection measurement $\{ \hat P( +;\bm n), \hat P(-;\bm n) \}$ if and only if $\varepsilon_1 = \varepsilon_2 =1$ and $r=1/2$. On the other hand,  the POVM is trivial (i.e., $\hat E_+ = r \hat I$ and $\hat E_- = (1-r) \hat I$) if and only if  $\varepsilon_1 = \varepsilon_2 = 0$; then we cannot obtain any information about $\hat \rho$.   Equations~(\ref{nonideal1}) can be rewritten as 
\begin{eqnarray}
\left( 
\begin{array}{c}
\hat E_+ \\
\hat E_- \\
\end{array} 
\right)
=F
\left( 
\begin{array}{c}
\hat P (+; \bm n) \\
\hat P (-; \bm n) \\
\end{array} 
\right),
\label{projection-povm}
\end{eqnarray}
where $F$ is the $2 \times 2$ transition-probability matrix
\begin{eqnarray}
F =
\left( 
\begin{array}{cc}
r (1+ \varepsilon_1) & r (1- \varepsilon_1) \\
(1-r) (1 - \varepsilon_2) & (1-r) (1+ \varepsilon_2) \\
\end{array} 
\right)
\label{transition}
\end{eqnarray}
which satisfies $\sum_i F_{ij} = 1$ and $0 \leq |\det F|^2 \leq 1$.
Note that $F$ describes a binary symmetric channel~\cite{Cover-Thomas} if and only if $r=1/2$ and $\varepsilon_1 = \varepsilon_2$.  It follows from Eq.~(\ref{projection-povm}) that any measurement process described by a POVM consisting of two positive operators is  formally equivalent to a measurement process in which a classical error is added to the projection measurement.  The physical origin of this error, however, lies in the quantum-mechanical interaction.

\textbf{Example 3} (\textit{probabilistic measurement}).
Suppose that a nonideal measurement of $\hat A = \bm n_A \cdot \hat{\bm \sigma}$ is performed with probability $\xi$ ($0<\xi<1$) and that $\hat B = \bm n_B \cdot \hat{\bm \sigma}$ is performed with probability $1- \xi$. The POVM corresponding to this probabilistic  measurement consists of four operators:
\begin{equation}
\textbf{E} = \{ \xi \hat E(\pm ;\bm n_A), (1-\xi )\hat E(\pm ;\bm n_B) \}.
\end{equation}
As the number of measured samples increases, this measurement  asymptotically approaches the measurements on $N$ identically prepared samples which are divided into two groups in the ratio $\xi :1-\xi$, with $\hat A$ being measured for the first group and $\hat B$ for the second group.

Other important examples such as nonideal joint measurements and quantum state tomography are discussed in Sec. V\hspace{-.1em}I.

\section{Accuracy Matrix}

\subsection{Definition of the Accuracy Matrix}

We will characterize the accuracy of an arbitrary observable in such a manner that it depends only on the process of measurement and  not on the measured state $\hat{\rho}$.  We first define the accuracy matrix.

\textbf{Definition 1} (\textit{accuracy matrix}).
The $3 \times 3$ accuracy matrix $\chi (\textbf{E})$  characterizing the measurement accuracy of  observables in terms of the POVM $\textbf{E}$  is defined as
\begin{equation}
\chi (\textbf{E})_{ij}\equiv \sum_k r_k (\bm v_k)_i (\bm v_k)_j,
\label{AM1}
\end{equation}
where $(\bm v_k)_i$ denotes the $i$th component of the real vector $\bm v_k$ and $ij$ shows indices of matrix elements of $\chi (\textbf{E})$.
We introduce the notation $\bm v  \bm v^{\rm T}$ with $\bm v \in \mathbb{R}^3$ as
\begin{equation}
(\bm v  \bm v^{\rm T})_{ij} \equiv (\bm v)_i (\bm v)_j;
\end{equation}
that is, $\bm v^{\rm T}$ denotes the transposed vector of $\bm v$ and $\bm v  \bm v^{\rm T}$ denotes  the projection matrix onto direction $\bm v$ in $\mathbb{R}^3$ whose $ij$ matrix element is given by $(\bm v)_i (\bm v)_j$. We can then rewrite (\ref{AM1}) in matrix form as
\begin{equation}
\chi (\textbf{E}) \equiv \sum_k r_k \bm v_k \bm v_k^{\rm T}.
\label{AM2}
\end{equation}
Note that $\chi (\textbf{E})$ is positive semidefinite and Hermitian, and can therefore be diagonalized by an orthonormal transformation.

The physical meaning and useful properties of the accuracy matrix will be investigated subsequently, and its foundation from an information-theoretic point of view will be  established  in terms of the maximum-likelihood estimation of the probability distribution of observables  in Sec.~V\hspace{-.1em}I.  In fact, the accuracy matrix is closely related to  Fisher information matrix (\ref{Fisher1}) or (\ref{Fisher2}), although physical quantities such as the  measurement error can be directly derived from the accuracy matrix without resort to  Fisher information.

Noting that $\sum_k r_k |\bm v_k |^2 \leq \sum_k r_k =1$, we can obtain the following fundamental inequality which forms the basis of  trade-off relations to be discussed later.

\textbf{Theorem 1.} 
Three eigenvalues $\{ \chi_1 , \chi_2, \chi_3 \}$ of $\chi (\textbf{E})$ satisfy 
\begin{equation}
\chi_1 + \chi_2 + \chi_3 \leq 1,
\label{trade-off1}
\end{equation}
or equivalently,
\begin{equation}
{\rm Sp} (\chi(\textbf{E})) \leq 1,
\label{trade-off1'}
\end{equation}
where we denote the trace of the $3 \times 3$ matrix as ${\rm Sp}(\cdots)$ to reserve symbol ${\rm tr}(\cdots)$ for the trace of a quantum-mechanical $2 \times 2$ matrix.  The equality $\chi_1 + \chi_2 + \chi_3 = 1$, or ${\rm Sp} (\chi) =1$, holds if and only if $| \bm v_k | = 1$ for all $k$.

The following corollary follows from the positivity of $\chi (\textbf{E})$.

\textbf{Corollary 1.}
The accuracy matrix satisfies the following matrix inequality:
\begin{equation}
0 \leq \chi(\textbf{E}) \leq I_3,
\label{trade-off2}
\end{equation}
where $I_3$ is the $3 \times 3$ identity matrix, and $\chi (\textbf{E})\leq I_3$ means that all eigenvalues of $I_3 - \chi(\textbf{E})$ are nonnegative.

The following examples illustrate the physical meaning of the accuracy matrix.  

We first consider a nonideal quantum measurement (see also example 2 in Sec.~I\hspace{-.1em}I).  We can rewrite Eq.(\ref{nonideal1}) as 
\begin{equation}
\hat E_1 = r (\hat I + \bm v_1 \cdot \hat{\bm \sigma}), \ \hat E_2 = (1-r) (\hat I + \bm v_2 \cdot \hat{\bm \sigma}),
\end{equation}
where $\bm v_1 = \varepsilon_1 \bm n$ and $\bm v_2 = - \varepsilon_2 \bm n$.  The accuracy matrix can  then be represented by
\begin{equation}
\chi (\textbf{E}) \equiv r \bm v_1 \bm v_1^{\rm T} +  (1-r) \bm v_2  \bm v_2^{\rm T} = 
\chi_{11} \bm n \bm n^{\rm T},
\label{chi1}
\end{equation}
where  $\chi_{11}$ is the eigenvalue of $\chi$ corresponding to the eigenvector $\bm n$, and is given by
\begin{equation}
\chi_{11}=r | \bm v_1 |^2 + (1-r) | \bm v_2 |^2 = \varepsilon_1 \varepsilon_2.
\label{chi11}
\end{equation}
We can also write $\chi_{11}$ in terms of the transition-probability matrix introduced in Eq.~(\ref{transition}) as
\begin{equation}
\chi_{11} = \frac{|\det F|^2}{4r} + \frac{|\det F|^2}{4(1-r)} = \frac{| \det F|^2}{4r(1-r)}. 
\end{equation}
The accuracy parameter $\chi_{11}$ satisfies
\begin{equation}
0 \leq \chi_{11} \leq 1,
\end{equation}
where $\chi_{11} = 1$ holds if and only if $| \bm v_1 | = | \bm v_2 | =1$ and $r=1/2$; that is, $\textbf{E}$ describes the projection measurement of observable $\bm n \cdot \hat{\bm \sigma}$. Note that $\chi (\textbf{E}) = \bm n \bm n^{\rm T}$ holds in this case. On the other hand, $\chi_{11} = 0$ holds if and only if $| \bm v_1 | = | \bm v_2 | =0$. In this case, $\chi (\textbf{E}) = O$ holds, and  we cannot obtain any information about $\hat \rho$.   The nonzero eigenvalue $\chi_{11}$ thus characterizes the measurement accuracy of $\bm n \cdot \hat{\bm \sigma}$; the larger $\chi_{11}$, the more information we can extract about $\bm n \cdot \hat{\bm \sigma}$ from the measurement outcome.  These properties can be generalized for an arbitrary  POVM as shown below.

Another example is the probabilistic  measurement of two noncommuting observables (see example 3 in Sec.~I\hspace{-.1em}I).  We consider the nonideal measurement of $\hat A$  whose accuracy matrix is $\chi_A \bm n_A \bm n_A^{\rm T}$ and that of $\hat B$ whose accuracy matrix is $\chi_B \bm n_B \bm n_B^{\rm T}$.   
The accuracy matrix of the probabilistic measurement is given by
\begin{equation}
\chi (\textbf{E}) = \xi \chi_A \bm n_A \bm n_A^{\rm T} + (1-\xi) \chi_B \bm n_B \bm n_B^{\rm T}.
\label{accuracy1}
\end{equation}
This representation suggests that the measurement accuracy concerning $\hat A$ is degraded by a factor of $\xi$ compared with the single nonideal measurement of $\hat A$, because we cannot observe $\hat A$ with probability $1-\xi$.  A similar argument applies to $\hat B$ as well.  Equation~(\ref{accuracy1}) shows that $\chi (\textbf{E})$ is the linear combination of the accuracy matrices of   POVMs measuring  $\hat A$ and  $\hat B$, where the coefficients   $\xi$ and $1- \xi$  give the probabilities of measuring $\hat A$ and $\hat B$, respectively.

This can be generalized as follows.   Let us consider three POVMs:  $\textbf{E}'= \{ \hat E_1, \hat E_2, \cdots, \hat E_m \}$,  $\textbf{E}'' = \{ \hat E_{m+1}, \hat E_{m+2}, \cdots, \hat E_n \}$, and $\textbf{E} = \{ \xi \hat E_1,   \cdots , \xi \hat E_m, (1-\xi ) \hat E_{m+1}, \cdots, (1-\xi) \hat E_n \}$ with $0 < \xi < 1$. The POVM $\textbf{E}$ describes the probabilistic   measurement of $\textbf{E}'$ with probability $\xi$ and that of $\textbf{E}''$ with probability $1-\xi$.   According to the definition of the accuracy matrix, we have
\begin{equation}
\begin{split}
\chi (\textbf{E}) &= \sum_{k=1}^m (\xi r_k) \bm v_k \bm v_k^{\rm T} + \sum_{k=m+1}^n \{(1-\xi) r_k \} \bm v_k \bm v_k^{\rm T} \\
&= \xi \chi (\textbf{E}') + (1- \xi )\chi (\textbf{E}'').
\end{split}
\end{equation}
We thus obtain  the following theorem.

\textbf{Theorem 2 } (\textit{linearity}):
\begin{equation}
\chi (\textbf{E}) = \xi \chi (\textbf{E}') + (1- \xi) \chi (\textbf{E}''),
\end{equation}
or more symbolically,
\begin{equation}
\chi \left( \xi \textbf{E}' + (1-\xi) \textbf{E}'' \right) = \xi \chi (\textbf{E}') + (1- \xi) \chi (\textbf{E}'').
\end{equation}

Note that we can take as a scalar measure of the measurement accuracy the largest eigenvalue of the accuracy matrix which we denote as $\chi (\textbf{E})_{\rm max}$.  It satisfies $0 \leq \chi (\textbf{E})_{\rm max} \leq 1$, where $\chi (\textbf{E})_{\rm max} = 1$ holds if and only if $\textbf{E}$ describes the projection measurement of a particular direction  and $\chi (\textbf{E})_{\rm max} = 0$ if and only if the POVM is trivial: $\textbf{E} = \{ q_k \hat I \}$, where  $\hat I$ is  the identity operator and $q_k$ denotes the probability of finding outcome $k$, with $\sum_k q_k =1$.  We may alternatively choose the scalar measure to be ${\rm Sp} (\chi (\textbf{E}))$; it has the linear property from theorem 2 and satisfies $0 \leq {\rm Sp} (\chi (\textbf{E})) \leq 1$, where ${\rm Sp} (\chi (\textbf{E})) = 0$ if and only if the POVM is trivial.

%%%%%%%%%%%%%%%%%%%%%%%%%%%%%%%%%%%%%%%%%%%%%%%%%%%%%%%%%%%%%%%%%%%%%%%%%%%%%%%%%%%%%%%%%%%%%%%%%%%%%%%%%%

\subsection{Accuracy Parameter in a Specific Direction}

We next parametrize the measurement accuracy of a particular observable.  We denote the support of $\chi ({\textbf E})$ as $V(\textbf{E})$; that is,  $V(\textbf{E})$ is the subspace of $\mathbb R^3$ spanned by all eigenvectors of $\chi (\textbf{E})$ with nonzero eigenvalues. 

\textbf{Definition 2 } (\textit{measurement accuracy}).
The accuracy parameter $\chi (\bm n; \textbf{E})$ in  direction   $\bm n \in V(\textbf{E})$ is defined as 
\begin{equation}
\chi (\bm n; \textbf{E}) \equiv \frac{1}{\bm n \cdot (\chi (\textbf{E})^{-1}) \bm n},
\label{A-parameter}
\end{equation}
where $\chi(\textbf{E})^{-1}$ is assumed to act only on subspace $V(\textbf{E})$.  If $\bm n  \in\hspace{-.80em}/ V(\textbf{E})$, we set $\chi(\bm n; \textbf{E}) = 0$.

This definition is closely related to the Fisher information concerning a particular direction defined in Eq.~(\ref{Fisher3}).

\textbf{Definition 3} (\textit{measurement error})
The error parameter of the measurement in direction $\bm n$ is defined as  
\begin{equation}
\varepsilon (\bm n; \textbf{E}) \equiv  \frac{1}{\chi(\bm n; \textbf{E})} - 1 = \bm n \cdot (\chi(\textbf{E})^{-1}) \bm n  -1.
\label{E-parameter}
\end{equation}

The parameters $\chi (\bm n; \textbf{E})$ and $\varepsilon (\bm n; \textbf{E})$ satisfy the following inequalities.

\textbf{Theorem 3}:
\begin{equation}
\ 0 \leq \chi (\bm n; \textbf{E}) \leq 1,
\label{inequality1}
\end{equation}
\begin{equation}
0 \leq \varepsilon (\bm n; \textbf{E}) \leq \infty.
\label{inequality2}
\end{equation}
The equality $\chi(\bm n; \textbf{E}) = 1$, or equivalently $\varepsilon (\bm n; \textbf{E}) = 0$, holds if and only if the measurement described by $\textbf{E}$ is equivalent to a projection measurement in  direction $\bm n$.  In this case, the measurement involves no measurement error. The other limit of $\chi (\bm n; \textbf{E}) = 0$, or equivalently $\varepsilon (\bm n; \textbf{E}) = \infty$, holds if and only if $\bm n \in\hspace{-.88em}/ V(\textbf{E})$. In this case, we cannot obtain any information about direction $\bm n$ from the measurement.

\textbf{Proof }
Since $\chi(\textbf{E})$ commutes with the identity operator $I_3$, we can show that $I_3 \leq \chi (\textbf{E})^{-1}$ from inequality (\ref{trade-off2}) in corollary 1.  We thus obtain
\begin{equation}
1 = \bm n \cdot I_3 \bm n \leq \bm n \cdot (\chi (\textbf{E})^{-1}) \bm n.
\label{inequality3}
\end{equation}
Inequalities (\ref{inequality1}) and (\ref{inequality2}) are the direct consequences of this inequality. The condition that $\chi(\bm n; \textbf{E})=0$ and $\varepsilon(\bm n; \textbf{E})=\infty$ hold follows from the definitions of $\chi(\bm n; \textbf{E})$ and $\varepsilon(\bm n; \textbf{E})$.

We next show the condition that $\chi(\bm n; \textbf{E})=1$ and $\varepsilon(\bm n; \textbf{E})=0$ hold.  If $\textbf{E}$ is the projection measurement in direction $\bm n$, then $\chi (\bm n; \textbf{E}) = 1$. Conversely, from inequality~(\ref{trade-off1}), it can be shown that if $\chi(\bm n; \textbf{E}) = 1$ and $\varepsilon (\bm n;\textbf{E}) =0$ hold, then $\bm n$ is the eigenvector corresponding to eigenvalue $1$ and that the other two eigenvalues are $0$. It follows from the condition of equality ${\rm Sp} (\chi (\textbf{E}) )=1$ in theorem 1 that $\bm v_k = 1$ for all $k$.  Therefore, without loss of generality, we can write the POVM as  
\begin{eqnarray}
\hat E_1 = r_1 (\hat I+ \bm n \cdot \hat{\bm \sigma}),  \cdots,  \ \hat E_m = r_m (\hat I+ \bm n \cdot \hat{\bm \sigma}), \\
\hat E_{m+1} = r_{m+1} (\hat I -  \bm n \cdot \hat{\bm \sigma}),  \cdots, \ \hat E_n = r_n (\hat I -  \bm n \cdot \hat{\bm \sigma}),
\end{eqnarray}
where $\sum_{k=1}^m r_k = \sum_{k=m+1}^n r_k = 1/2$, because  $\sum_k r_k = 1$ and $\sum_{k=1}^m r_k \bm n - \sum_{k=m+1}^n r_k \bm n = \bm 0$ hold.  We define two operators as
\begin{equation}
\hat P(+; \bm n) \equiv \sum_{k=1}^m \hat E_k, \ \hat P(-; \bm n) \equiv \sum_{k=m+1}^n \hat E_k;
\end{equation}
then $\{  \hat P(+; \bm n), \hat P(-; \bm n) \}$ describes the projection measurement in direction $\bm n$.---

Let $\bm n_1$, $\bm n_2$, and $\bm n_3$ be the eigenvectors of $\chi (\textbf{E})$, and $\chi_1$, $\chi_2$, and $\chi_3$ be the corresponding eigenvalues.  It can be shown that
\begin{equation}
\chi (\bm n_i; \textbf{E}) = \chi_i, \ (i=1,2,3).
\end{equation}
According to theorem 1, we  cannot  simultaneously measure the three directions corresponding to the eigenvectors with the maximum accuracy $\chi_i = 1$  for all $i$; the trade-off relation~(\ref{trade-off1}) or (\ref{trade-off1'}) is equivalent to
\begin{equation}
\chi_1 + \chi_2 + \chi_3 \leq 1.
\label{trade-off1''''}
\end{equation}
This trade-off relation represents the uncertainty relation between the measurement errors in the three directions.

We define that the POVM $\textbf{E}$ is optimal if and only if ${\rm Sp} (\chi (\textbf{E}))= \chi_1 + \chi_2 + \chi_3 = 1$; that is, $\textbf{E}$ reaches the upper bound of trade-off relation~(\ref{trade-off1}), (\ref{trade-off1'}), or (\ref{trade-off1''''}).   On the other hand,  we define that $\textbf{E}$ is symmetric if and only if $\chi (\bm n; \textbf{E}) = \chi (\bm n'; \textbf{E})$ holds for any $\bm n$ and $\bm n'$.  In this case,  $\chi (\textbf{E})$ is proportional to the $3 \times 3$ identity matrix.

\subsection{Reconstructive subspace}

We next introduce the concept of ``reconstructive subspace'' and ``reconstructive direction.''  The following theorem can be directly shown from the definition of the accuracy matrix.

\textbf{Theorem 4 }
$V(\textbf{E})$ corresponds to the subspace spanned by the set of basis vectors $\{ \bm v_k \}$ of the accuracy matrix~(\ref{AM2}). 

Suppose that we perform the measurement $\{ \hat E_k \}$ and obtain the probability distribution $\{ q_k \}$ for each outcome $k$.  Can we then reconstruct the premeasurement distribution $\{ p(\bm n; \pm) \}$ of the system  from $\{ q_k \}$?  The answer is given by the following theorem.

\textbf{Theorem 5} (\textit{reconstructive subspace and reconstructive direction})
We can reconstruct  the probability distribution $\{ p(\pm ;\bm n) \}$ from the measured distribution $\{ q_k \}$ if and only if $\bm n \in V(\textbf{E})$.  We thus refer to  $V(\textbf{E})$ as a reconstructive subspace  and to a unit vector in $V(\textbf{E})$ as a reconstructive direction. 

\textbf{Proof }
We can show from Eq.~(\ref{qk}) that
\begin{equation}
\left( 
\begin{array}{c}
q_1 \\
q_2 \\
\vdots \\
q_m
\end{array} 
\right)
= M 
\left( 
\begin{array}{c}
(\bm s_0)_x \\
(\bm s_0)_y \\
(\bm s_0)_z
\end{array} 
\right)
+
\left( 
\begin{array}{c}
r_1 \\
r_2 \\
\vdots \\
r_m
\end{array} 
\right),
\end{equation}
where $M$ is a $m \times 3$ matrix:
\begin{equation}
M = 
\left( 
\begin{array}{ccc}
r_1 (\bm v_1)_x  & r_1 (\bm v_1)_y & r_1 (\bm v_1)_z \\
r_2 (\bm v_2)_x & r_2 (\bm v_2)_y & r_2 (\bm v_2)_z \\
\ & \vdots & \ \\
r_m (\bm v_m)_x  & r_m (\bm v_m)_y & r_m (\bm v_m)_z
\end{array} 
\right).
\end{equation}
Let ${\rm Ker} (M)$ and ${\rm Im}(M)$ be the kernel and  image of $M$, respectively. It can  easily be shown that ${\rm Ker} (M) = V(\textbf{E})^{\perp}$.  Let us introduce the equivalence relation ``$\sim$''  as $\bm a \sim \bm b \  \Leftrightarrow \ \bm a - \bm b \in  V(\textbf{E})^{\perp}$. We denote the equivalence class of $\bm v \in \mathbb R^3$ as $ [ \bm v ] $, where $[\bm v]$ is an element of the quotient space $\mathbb R^3 / \sim$.  From the homomorphism theorem, the quotient map $M /\sim$ is a linear isomorphism from $\mathbb R^3 / \sim$ to ${\rm Im} (M)$. Noting that $(q_1, q_2, \cdots q_m)^t - (r_1, r_2, \cdots r_m)^t \in {\rm Im} (M)$, we obtain
\begin{equation}
[ \bm s_0 ]
=
( M/ \sim )^{-1}
\left(
\left( 
\begin{array}{c}
q_1 \\
q_2 \\
\vdots \\
q_m
\end{array} 
\right)
-
\left( 
\begin{array}{c}
r_1 \\
r_2 \\
\vdots \\
r_m
\end{array} 
\right)
\right).
\end{equation}
By taking a representative $\bm s_0' \in [\bm s_0] $, we can  reconstruct $\bm n \cdot \bm s_0$ as $\bm n \cdot \bm s_0 = \bm n \cdot \bm s_0'$ which gives $p(\pm;\bm n)$ through Eq.~(\ref{distribution}).---

We consider the nonideal measurement  of $\hat A = \bm n_A \cdot \hat{\bm \sigma}$ with the POVM $\textbf{E}_A = \{ \hat E (+; \bm n_A), \hat E (-;\bm n_A) \}$ in Eq.~(\ref{nonideal1}).   The  nonideal measurement is characterized with the  accuracy matrix  $\chi_A \bm n_A \bm n_A^{\rm T}$.  In this case, we can show that $V(\textbf{E}_A) = \{ \bm a \ | \ \bm a = \lambda \bm n_A, \lambda \in \textbf{R} \}$. It follows that $\chi(\pm \bm n_A ; \textbf{E}) = \chi_A$ and $\chi (\bm n ; \textbf{E}) = 0$ for $\bm n \neq \pm \bm n_A $.

We next consider the probabilistic measurement of $\textbf{E}_A$ and $\textbf{E}_B$ in example 3 in Sec.~I\hspace{-.1em}I.  The probabilistic measurement is characterized with the  accuracy matrix of the  joint POVM $\textbf{E}$ given in (\ref{accuracy1}), so the reconstructive subspace is two dimensional: $V(\textbf{E}) = \{ \lambda_A \bm n_A + \lambda_B \bm n_B \ | \ (\lambda_A, \lambda_B) \in \mathbb R^2 \}$. A straightforward  calculation shows that 
\begin{equation}
\chi(\bm n_A; \textbf{E}) = \xi \chi_A, \ \chi (\bm n_B ; \textbf{E}) = (1-\xi) \chi_B.
\label{accuracy2}
\end{equation}

%%%%%%%%%%%%%%%%%%%%%%%%%%%%%%%%%%%%%%%%%%%%%%%%%%%%%%%%%%%%%%%%%%%%%%%%%%%%%%%%%%%%%%%%%%%%%%%%%%%%%%%%%%%%%%

\subsection{Data Processing Inequality}

If the classical noise described by a transition-probability matrix is added to the measurement outcomes,  the measurement accuracy should deteriorate.  This fact can be expressed as a data processing inequality.

\textbf{Theorem 6 } (\textit{data processing inequality}).
Suppose that two POVMs  $\textbf{E} = \{ \hat E_1, \hat E_2 , \cdots , \hat E_n \}$ and  $\textbf{E}' = \{ \hat E_1' , \hat E_2' , \cdots, \hat E_m' \}$ are related to each other by
\begin{equation}
\hat E_j' = \sum_{k=1}^n F_{jk} \hat E_k,
\label{data-processing2}
\end{equation}
where $F_{ij}$ is an $m \times n$ transition-probability matrix satisfying $\sum_j F_{jk} =1$. It then follows that
\begin{equation}
\chi (\textbf{E}') \leq \chi(\textbf{E}),
\label{data-processing}
\end{equation}
where matrix inequality~(\ref{data-processing}) means that all the eigenvalues of $\chi(\textbf{E})-\chi(\textbf{E}')$ are non-negative.

\textbf{Proof }
We can parametrize the POVMs as
\begin{equation}
\hat E_k =  r_k (\hat I + \bm v_k \cdot \hat{\bm \sigma}),
\end{equation}
\begin{equation}
\hat E_j' = \left( \sum_{k=1}^n \xi_{jk} \right) \left( \hat I + \frac{\sum_{k=1}^n \xi_{jk} \bm v_k}{\sum_{k=1}^n \xi_{jk}} \cdot \hat{\bm \sigma} \right),
\end{equation}
where $\xi_{jk} \equiv F_{jk}r_k$. Introducing the function $f_{\bm w}; \ \mathbb R^3 \to \mathbb R^3$, with arbitrary vector $\bm w \in \mathbb R^3$ as
\begin{equation}
f_{\bm w} (\bm v) \equiv  \bm w \cdot (\bm v \bm v^{\rm T}) \bm w = |\bm v \cdot \bm w |^2, 
\end{equation}
we can show that
\begin{equation}
\bm w \cdot \chi (\textbf{E}) \bm w = \sum_{k=1}^n r_k f_{\bm w} (\bm v_k),
\end{equation}
\begin{equation}
\bm w \cdot \chi (\textbf{E}') \bm w = \sum_{j=1}^m \left( \sum_{k=1}^n \xi_{jk} \right) f_{\bm w} \left( \frac{\sum_{k=1}^n \xi_{jk} \bm v_k}{\sum_{k=1}^n \xi_{jk}} \right).
\end{equation}
The Hessian of $f_{\bm w}$, which is defined as $H(f_{\bm w})_{ij} \equiv \partial^2 f_{\bm w} (\bm v) / \partial (\bm v)_i \partial (\bm v)_j$, becomes
\begin{equation}
H(f_{\bm w}) = \bm w  \bm w^{\rm T} \geq 0,
\end{equation}
so that $f_{\bm w}$ is a concave function. Therefore 
\begin{equation}
f_{\bm w} \left( \sum_k \xi_k \bm v_k \right) \leq \sum_k \xi_k f_{\bm w} (\bm v_k)
\label{con2}
\end{equation}
holds for any $\{ \xi_k \}$ satisfying $\sum_k \xi_k =1$ and $0 \leq \xi_k \leq 1$ for all $k$. Taking $\xi_k = \xi_{jk} / \sum_{k'=1}^n \xi_{jk'}$, inequality~(\ref{con2}) becomes
\begin{equation}
f_{\bm w} \left( \frac{\sum_{k=1}^n \xi_{jk} \bm v_k}{\sum_{k=1}^m \xi_{jk}} \right) \leq \frac{\sum_{k=1}^n \xi_{jk}  f_{\bm w} (\bm v_k)}{\sum_{k=1}^n \xi_{jk}},
\end{equation}
or equivalently,
\begin{equation}
\left( \sum_{k=1}^n \xi_{jk} \right) f_{\bm w} \left( \frac{\sum_{k=1}^m \xi_{jk} \bm v_k}{\sum_{k=1}^n \xi_{jk}} \right) \leq \sum_{k=1}^n \xi_{jk}  f_{\bm w} (\bm v_k).
\end{equation}
Noting that $\sum_{j=1}^m \xi_{jk} =r_k$, we obtain
\begin{equation}
\sum_{j=1}^m \left( \sum_{k=1}^n \xi_{jk} \right) f_{\bm w} \left( \frac{\sum_{k=1}^n \xi_{jk} \bm v_k}{\sum_{k=1}^n \xi_{jk}} \right) \leq  \sum_{k=1}^n r_k f_{\bm w} (\bm v_k),
\end{equation}
which implies that
\begin{equation}
\bm w \cdot \chi (\textbf{E}') \bm w \leq \bm w \cdot \chi (\textbf{E}) \bm w.
\label{52}
\end{equation}
Since~(\ref{52}) holds for arbitrary $\bm w$, we   obtain  (\ref{data-processing}).---

The following corollary is a direct consequence of the foregoing theorem.

\textbf{Corollary 6}
Suppose that $\textbf{E}' =\{  \hat E_1', \hat E_2' , \cdots , \hat E_m' \}$ is obtained by a coarse graining of $\textbf{E} = \{ \hat E_1, \hat E_2 , \cdots , \hat E_n \}$: $\hat E_1' = \hat E_1 + \hat E_2 + \cdots + \hat E_{i(1)}, \ \hat E_2' = \hat E_{i(1) +1} + \cdots + \hat E_{i(2)} , \ \cdots$, and $\hat E_m' = \hat E_{i(m-1)+1} + \cdots  + \hat E_n$, with $1 < i(1) < i(2) < \cdots < i(m-1) < n$. Then 
\begin{equation}
\chi (\textbf{E}') \leq \chi (\textbf{E})
\label{grain}
\end{equation}
holds.  Inequality~(\ref{grain}) means that the measurement accuracy in any  direction is decreased by a coarse graining. 

We can also express the data processing inequality in terms of the accuracy parameter in an arbitrary  direction.

\textbf{Theorem 7 }
(\textit{data processing inequality}).
We consider the POVMs $\textbf{E}$ and $\textbf{E}'$ satisfying Eq.~(\ref{data-processing2}).  Suppose that $V(\textbf{E})=V(\textbf{E}')=\mathbb R^3$.  Then,
\begin{equation}
\chi(\textbf{E}')^{-1} \geq \chi(\textbf{E})^{-1}
\label{data-processing3}
\end{equation}
holds, or equivalently, 
\begin{equation}
\chi (\bm n; \textbf{E}') \leq \chi (\bm n; \textbf{E})
\label{data-processing4}
\end{equation}
holds for arbitrary $\bm n$.

\textbf{Proof }
Let $\chi_1$, $\chi_2$, and $\chi_3$ be the eigenvalues of $\chi(\textbf{E})$, and $\bm n_1$, $\bm n_2$, and $\bm n_3$ be the corresponding eigenvectors.  Similarly, let $\chi_1'$, $\chi_2'$, and $\chi_3'$ be the eigenvalues of $\chi(\textbf{E}')$, and $\bm n_1'$, $\bm n_2'$, and $\bm n_3'$ be the corresponding eigenvectors.  It follows from the data processing inequality~(\ref{data-processing})   that
\begin{equation}
\bm n_i \cdot \chi(\textbf{E}') \bm n_i = \sum_{j=1}^3 n_{ij}^2 \chi_j' \leq \chi_i = \bm n_i \cdot \chi(\textbf{E}) \bm n_i, 
\end{equation} 
for $i=1,2,3$, where $n_{ij} \equiv \bm n_i \cdot \bm n_j'$.  Applying the concave inequality to $1/x$, we obtain
\begin{equation}
\chi_i^{-1} \leq \left(  \sum_{j=1}^3 n_{ij} \chi_j' \right)^{-1} \leq \sum_{j=1}^3 n_{ij}^2 \chi_j'^{-1}.
\end{equation}
For  arbitrary $\bm n$, we can show that
\begin{equation}
\begin{split}
\bm n \cdot (\chi(\textbf{E})^{-1}) \bm n &= \sum_{i=1}^3 (\bm n \cdot \bm n_i)^2 \chi_i^{-1} \\
&\leq \sum_{i,j=1}^3 (\bm n \cdot \bm n_i)^2 n_{ij}^2 \chi_j'^{-1} \\
&= \sum_{j=1}^3  (\bm n \cdot \bm n_i')^2 \chi_j'^{-1} \\
&= \bm n \cdot (\chi(\textbf{E}')^{-1}) \bm n,
\end{split}
\end{equation}
which implies (\ref{data-processing3}) and (\ref{data-processing4}).---

%%%%%%%%%%%%%%%%%%%%%%%%%%%%%%%%%%%%%%%%%%%%%%%%%%%%%%%%%%%%%%%%%%%%%%%%%%%%%%%%%%%%%%%%%%%%%%%%%%%%%%%%%%

\section{Trade-off Relations for Generalized Simultaneous Measurement of a Qubit System}

We now derive  general trade-off relations between the measurement errors of noncommuting observables, which are the main results of this paper.

Let $\bm n_1$, $\bm n_2$, and $\bm n_3$ be the respective eigenvectors of $\chi (\textbf{E})$ corresponding to the eigenvalues $\chi_1$, $\chi_2$, and $\chi_3$, where $\chi_i = \chi (\bm n_i;\textbf{E})$ ($i=1,2,3$).  We define the error parameters as  $\varepsilon_i \equiv \varepsilon (\bm n_i;\textbf{E}) = (\chi_i)^{-1} - 1$.  Inequality (\ref{trade-off1}) or (\ref{trade-off1'}) in theorem 1 can be rewritten in terms of the error parameters as
\begin{equation}
\varepsilon_1 \varepsilon_2 \varepsilon_3 \geq \varepsilon_1 + \varepsilon_2 + \varepsilon_3 + 2.
\label{trade-off1''}
\end{equation}
Considering  two eigenvalues alone (i.e., $\chi_1 + \chi_2 \leq 1$), we can simplify the trade-off relation:
\begin{equation}
\varepsilon_1 \varepsilon_2 \geq 1.
\label{trade-off1'''}
\end{equation}

The trade-off relations (\ref{trade-off1''}) and (\ref{trade-off1'''}) can be generalized to the case of arbitrary directions.  We first consider the case of  two observables.

\textbf{Theorem 8 } (\textit{trade-off relation}).
We consider a simultaneous measurement in two directions $\bm n_A$ and $\bm n_B$ ($\bm n_A \cdot \bm n_B = \cos \theta$) described  by the POVM $\textbf{E}$.  We assume $\bm n_A \in V(\textbf{E})$ and $\bm n_B \in V(\textbf{E})$, and define $\varepsilon_{\alpha} \equiv \varepsilon (\bm n_{\alpha};\textbf{E})$ and $\chi_{\alpha} \equiv \chi (\bm n_{\alpha};\textbf{E})$ ($\alpha = A,B$). Then the trade-off relation 
\begin{equation}
\varepsilon_A \varepsilon_B \geq \sin^2 \theta,
\label{trade-off3}
\end{equation}
or equivalently,
\begin{equation}
\chi_A + \chi_B - \chi_A \chi_B \cos^2 \theta \leq 1
\label{trade-off4}
\end{equation}
holds. 

\textbf{Proof.}
We divide the proof into two steps.

\textit{Step 1.}
We consider a situation in which both $\bm n_A$ and $\bm n_B$ lie in a plane spanned by two eigenvectors. Without loss of generality, we choose $\bm n_1$ and $\bm n_2$ as the two eigenvectors, and  expand $\bm n_A$ and $\bm n_B$ as
\begin{equation}
\bm n_A = \bm n_1 \cos \theta_A + \bm n_2 \sin \theta_A, 
\end{equation}
\begin{equation}
\bm n_B = \bm n_1 \cos \theta_B + \bm n_2 \sin \theta_B,
\end{equation}
where $\theta = \theta_A - \theta_B$.  It can be shown that
\begin{equation}
\varepsilon_A = \varepsilon_1 \cos^2 \theta_A + \varepsilon_2 \sin^2 \theta_A,
\end{equation}
\begin{equation}
\varepsilon_B = \varepsilon_1 \cos^2 \theta_B + \varepsilon_2 \sin^2 \theta_B.
\end{equation}
Applying the Cauchy-Schwarz inequality and $\varepsilon_1 \varepsilon_2 \geq 1$, we obtain
\begin{eqnarray}
\varepsilon_A \varepsilon_B &\geq& \left(  \sqrt{\varepsilon_1 \varepsilon_2} (\cos \theta_A \sin \theta_B - \sin \theta_A \cos \theta_B) \right)^2 \label{proof-1} \\
&=& \varepsilon_1 \varepsilon_2 \sin^2 \theta \\
&\geq& \sin^2 \theta. \label{proof-2}
\end{eqnarray}
The equality $\varepsilon_A \varepsilon_B = \sin^2 \theta$ holds if and only if $\varepsilon_1 \cos \theta_A \cos \theta_B + \varepsilon_2 \sin \theta_A \sin \theta_B = 0$ and $\varepsilon_1 \varepsilon_2=1$.  In the case of $\varepsilon_A = \varepsilon_B$ (i.e., the measurement errors are symmetric), the equality holds if and only if $\sin (\theta_A+\theta_B)\sin (\theta_A-\theta_B) \cos (\theta_A+\theta_B) = 0$.

\textit{Step 2.}
We next consider a more general case.  We choose an orthonormal basis $\{ \bm n'_1, \bm n'_2, \bm n'_3 \}$ such that both $\bm n_A$ and $\bm n_B$ are in the plane spanned by $\bm n'_1$ and $\bm n'_2$. We introduce the notation $\varepsilon'_i \equiv \varepsilon (\bm n'_i;\textbf{E})$ ($i=1,2,3$).  Let $Q_{ij}$ be a $3 \times 3$ orthogonal matrix which transforms $\{ \bm n_1, \bm n_2, \bm n_3 \}$ into $\{ \bm n'_1, \bm n'_2, \bm n'_3 \}$.  It can be shown that
\begin{equation}
\varepsilon_i' = \sum_{j=1}^3 Q_{ij}^2 \varepsilon_j.
\end{equation}
Note that $\sum_j Q_{ij}^2 =1$  because $Q_{ij}$ is an orthogonal matrix, and that the function $(1+x)^{-1}$ is concave.  It follows from a concave inequality  that
\begin{equation}
\frac{1}{\varepsilon_i' +1} \leq \sum_{j=1}^3 \frac{Q_{ij}^2}{\varepsilon_j +1}.
\end{equation}
Combining this with $\sum_i Q_{ij}^2 = 1$, we obtain
\begin{equation}
\sum_{i=1}^3 \frac{1}{\varepsilon_i' + 1} \leq \sum_{i=1}^3 \sum_{j=1}^3 \frac{Q_{ij}^2}{\varepsilon_j +1} =  \sum _{j=1}^3 \frac{1}{\varepsilon_j +1}.
\end{equation}
Therefore
\begin{equation}
\frac{1}{\varepsilon_1' + 1} + \frac{1}{\varepsilon_2' +1} \leq \sum_{i=1}^3 \frac{1}{\varepsilon_i' + 1} \leq \sum _{j=1}^3 \frac{1}{\varepsilon_j +1} \leq 1.
\end{equation}
This inequality means that $\chi'_1 + \chi'_2 \leq 1$, or equivalently, 
\begin{equation}
\varepsilon_1' \varepsilon_2' \geq 1.
\end{equation}
We can derive inequality (\ref{trade-off3}) by following the same procedure as in step 1.  We can directly derive inequality (\ref{trade-off4}) from (\ref{trade-off3}).---

We note that the equalities in~(\ref{trade-off3}) and~(\ref{trade-off4}) hold in the case that the POVM $\textbf{E}$ is given by  $\{ |\bm x_i | \hat I \pm  \bm x_i \cdot \hat{\bm \sigma} \}$ ($i=1,2$), where $\bm x_1 \equiv (\sqrt{\chi_A}\bm n_A + \sqrt{\chi_B} \bm n_B)/4$ and  $\bm x_2 \equiv (\sqrt{\chi_A}\bm n_A - \sqrt{\chi_B} \bm n_B)/4$.

The accessible regime for $\chi_A$ and $\chi_B$ is illustrated in Fig.1 for the case of $\theta = \pi/2$, $\theta = \pi / 6$, and $\theta = 0$.  Note that regime Q can be reached only through simultaneous measurement for the case of $\theta=\pi/6$~\cite{Kurotani}.

\begin{figure}[htbp]
 \begin{center}
  \includegraphics[width=70mm]{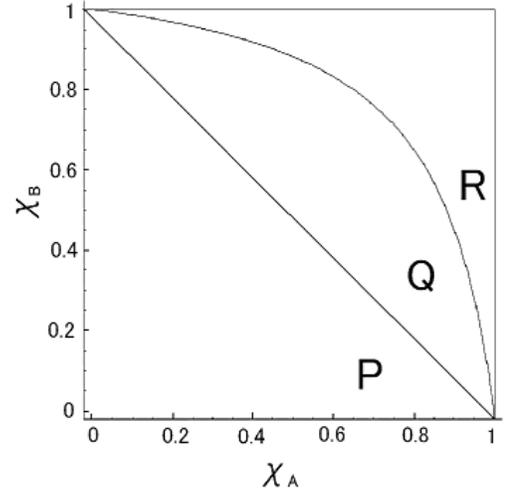}
 \end{center}
 \caption{ Trade-off relation for the accuracy of noncommuting observables. P indicates the regimes satisfying the inequality for the case of $\theta=\pi/2$, the union of P and Q indicate the regime satisfying inequality~(\ref{trade-off4}) for the case of $\theta=\pi/6$,  and the union of P, Q  and R indicate the regime satisfying the inequality for the case of $\theta=0$. We can only access regime Q through  simultaneous measurement for the case of $\theta=\pi/6$.}
 \label{fig:Trade-off}
\end{figure}

The trade-off relation can be interpreted as the  uncertainty relation between measurement errors~\cite{Neumann,Bohr,Arthurs-Kelly,Busch,Yamamoto-Haus,Arthurs-Goodman,Martens-Muynck,Appleby,Muynck,Busch-Shilladay,Hall,Ozawa-1,Andersson-Barnett-Aspect,Massar,Kurotani,Jammer}. It offers a rigorous representation of Bohr's principle of complementarity~\cite{Bohr} which dictates  ``the mutual exclusion of any two experimental procedures'' when we measure two noncommuting observables simultaneously.

The trade-off relation  between three observables can be formulated as follows.

\textbf{Theorem 9 }
We consider a simultaneous measurement in three directions $\bm n_A$, $\bm n_B$, and $\bm n_C$ described by the POVM $\textbf{E}$.  Let us assume that $\bm n_A$, $\bm n_B$, and $\bm n_C$ are linearly independent.  We set the notation $\varepsilon_{\alpha} \equiv \varepsilon (\bm n_{\alpha}; \textbf{E})$ and $\chi_{\alpha} \equiv \chi (\bm n_{\alpha}; \textbf{E})$, where $\alpha = A,B,C$. Then the inequality 
\begin{equation}
\varepsilon_A \varepsilon_B \varepsilon_C \geq 8 \{ \bm n_A \cdot (\bm n_B \times \bm n_C) \}^2
\label{trade-off5}
\end{equation}
holds.  The equality in (\ref{trade-off5}) holds if and only if $\varepsilon_1 = \varepsilon_2 = \varepsilon_3=2$ and $\{ \bm n_\alpha \}$ are orthogonal.

\textbf{Proof }
Introducing the notation
\begin{equation}
\tilde{\bm n}_\alpha \equiv 
\left( 
\begin{array}{c}
\sqrt{\varepsilon_1} (\bm n_\alpha)_1 \\
\sqrt{\varepsilon_2} (\bm n_\alpha)_2 \\
\sqrt{\varepsilon_3} (\bm n_\alpha)_3
\end{array} 
\right),
\end{equation}
where $(\bm n_\alpha)_i \equiv \bm n_\alpha \cdot \bm n_i$ ($\alpha = A,B,C, \ i=1,2,3$), it can be shown that
\begin{equation}
\varepsilon_\alpha = | \tilde{\bm n}_\alpha |^2.
\end{equation}
We thus obtain
\begin{equation}
\begin{split}
\varepsilon_A \varepsilon_B \varepsilon_C &= | \tilde{\bm n}_A |^2 | \tilde{\bm n}_B |^2 | \tilde{\bm n}_C |^2 \\
&\geq \{ \tilde{\bm n}_A \cdot (\tilde{\bm n}_B \times \tilde{\bm n}_C) \}^2 \\
&= \varepsilon_1 \varepsilon_2 \varepsilon_3  \{ \bm n_A \cdot (\bm n_B \times \bm n_C) \}^2 \\
&\geq (\varepsilon_1 + \varepsilon_2 + \varepsilon_3 + 2 )  \{ \bm n_A \cdot (\bm n_B \times \bm n_C) \}^2.
\end{split}
\end{equation}
We can show $\varepsilon_1 + \varepsilon_2 + \varepsilon_3 \geq 6$ from $\chi_1 + \chi_2 + \chi_3 \leq 1$; therefore we obtain (\ref{trade-off5}).---

%%%%%%%%%%%%%%%%%%%%%%%%%%%%%%%%%%%%%%%%%%%%%%%%%%%%%%%%%%%%%%%%%%%%%%%%%%%%%%%%%%%%%%%%%%%%%%%%%%%%%%%%%%%%%%%%%%

\section{Applications}

We have discussed in Sec.~I\hspace{-.1em}V trade-off relations (\ref{trade-off3}) and (\ref{trade-off4}) which describe the uncertainty relations in generalized simultaneous measurements.  In this section, we discuss possible applications of  these trade-off relations.

%%%%%%%%%%%%%%%%%%%%%%%%%%%%%%%%%%%%%%%%%%%%%%%%%%%%%%%%%%%%%%%%%%%%%%%%%%%%%%%%%%%%%%%%%%%%%%%%%%%%%%%%%%%%%%%

\subsection{Nonideal Joint Measurement}

We consider a class of simultaneous measurements called  nonideal joint measurements, where  two observables $\hat A = \bm n_A \cdot \hat{\bm \sigma}$ and $\hat B = \bm n_B \cdot \hat{\bm \sigma}$ are simultaneously measured.   Since their eigenvalues are $\pm 1$, each measurement should give a pair of outcomes $(i,j)$ ($i,j=\pm$) for observables $\hat A$ and $\hat B$. The joint POVM $\textbf{E} = \{ \hat E (i,j) \}$ can be parametrized as
\begin{equation}
\hat E (i,j) = r_{ij} (\hat I + \bm v_{ij} \cdot \hat{\bm \sigma}).
\label{joint1}
\end{equation}
The marginal POVMs $\textbf{E}_\alpha = \{ \hat E_\alpha (i) \}$ ($\alpha = A,B$)  are defined by
\begin{equation}
\begin{split}
\hat E_A(+) \equiv \hat E(+,+) + \hat E(+,-), \ \hat{E}_A(-) \equiv \hat{I} - \hat{E}_A(+), \\
\hat E_B(+) \equiv \hat E(+,+) + \hat E(-,+), \ \hat{E}_B(-) \equiv \hat{I} - \hat{E}_B(+),
\end{split}
\end{equation}
and can be parametrized by
\begin{equation}
\hat{E}_{\alpha}(+) = r_{\alpha} (\hat{I} + \bm{v}_{\alpha} \cdot \hat{\bm{\sigma}}), 
\end{equation}
where
\begin{equation}
\begin{split}
r_A = r_{++}+r_{+-}, \ \bm v_A = \frac{r_{++}\bm v_{++} + r_{+-}\bm v_{+-}}{r_{++}+r_{+-}}, \\
r_B = r_{++}+r_{-+}, \ \bm v_B = \frac{r_{++}\bm v_{++} + r_{-+}\bm v_{-+}}{r_{++}+r_{-+}}.
\end{split}
\end{equation}
This simultaneous measurement can be regarded as a nonideal joint measurement~\cite{Martens-Muynck,Muynck} if and only if the marginal  POVM $\textbf{E}_{A(B)}$ corresponds to the nonideal measurement of $\hat A (\hat B)$, that is,
\begin{equation}
\bm v_{\alpha} \| \bm n_{\alpha}.
\label{joint2}
\end{equation}
We can define the $2\times 2$ transition-probability matrices of $\textbf{E}_\alpha$ as in Eq.~(\ref{transition}):
\begin{eqnarray}
F_\alpha =
\left( 
\begin{array}{cc}
r_\alpha (1\pm | \bm v_\alpha |) & r_\alpha (1\mp | \bm v_\alpha |) \\
1-r_\alpha (1\pm | \bm v_\alpha |) & 1- r_\alpha (1\mp | \bm v_\alpha |) \\
\end{array} 
\right).
\end{eqnarray}

In this case, we can calculate the accuracy of $\hat A$ and $\hat B$ by two different methods.  One  method to calculate the accuracy parameter is based on the joint POVM $\textbf{E}$:
\begin{equation}
\chi_\alpha \equiv \chi (\bm n_\alpha ; \textbf{E}),
\end{equation}
where $\alpha = A,B$.  The other is based on the  marginal POVM $\textbf{E}_\alpha$:
\begin{equation}
\chi_\alpha' \equiv \chi (\bm n_\alpha; \textbf{E}_\alpha).
\end{equation}
Note that  $\chi (\textbf{E}_\alpha) = \chi_\alpha' \bm n_\alpha  \bm n_\alpha^{\rm T}$.

These two accuracy parameters are equivalent as shown by the following theorem.

\textbf{Theorem 10}:
\begin{equation}
\chi_{\alpha} = \chi'_{\alpha}.
\end{equation}

The proof of theorem 10 is   given in the Appendix.

For $\bm n \neq \pm \bm n_\alpha$,  we can show that $\bm n$ is not an element of $V(\textbf{E}_\alpha)$; therefore,  $\chi (\bm n; \textbf{E}_\alpha)=0$.  On the other hand, $\chi (\bm n; \textbf{E}) \geq 0$ holds by definition. We  can thus obtain the following corollary.

\textbf{Corollary 10 }
For arbitrary $\bm n$, 
\begin{equation}
\chi (\bm n ; \textbf{E}_\alpha) \leq \chi (\bm n ; \textbf{E}).
\label{data-processing5}
\end{equation}

We next discuss the relationship between the present work and our earlier work~\cite{Kurotani} for the case of nonideal joint measurement.   In Ref.~\cite{Kurotani}, we have introduced the accuracy parameter ${\mathcal X}_\alpha$ and error parameter  ${\mathcal E}_\alpha$ as
\begin{equation}
{\mathcal X}_\alpha \equiv (\det F_\alpha )^2, \ {\mathcal E}_\alpha \equiv \frac{1}{{\mathcal X}_\alpha} - 1.
\end{equation} 
On the other hand,  the accuracy parameter $\chi_\alpha'$ and  error parameter $\varepsilon_\alpha$ in the present paper are given by
\begin{equation}
\chi_\alpha = \chi_\alpha' = \frac{(\det F_\alpha )^2}{4 r_\alpha(1-r_\alpha)}, \ \varepsilon_\alpha = \frac{1}{\chi_\alpha} - 1.
\end{equation}
It can be easily shown that
\begin{equation}
\chi_\alpha \geq {\mathcal X}_\alpha, \ \varepsilon_\alpha \leq {\mathcal E}_\alpha,
\end{equation}
so the trade-off relations derived in the present paper  are  stronger than our previous ones (${\mathcal E}_A{\mathcal E}_B \geq \sin^2 \theta$  and ${\mathcal X}_A + {\mathcal X}_B - {\mathcal X}_A{\mathcal X}_B\cos^2 \theta \leq 1$) derived in Ref.~\cite{Kurotani}.
The latter trade-off relations can thus  be derived from those obtained in the present paper.

%%%%%%%%%%%%%%%%%%%%%%%%%%%%%%%%%%%%%%%%%%%%%%%%%%%%%%%%%%%%%%%%%%%%%%%%%%%%%%%%%%%%%%%%%%%%%%%%%%%%%%%%%%%%%%5%%%%

\subsection{Uncertainty Relation between Measurement Error and Back-action}

We have interpreted trade-off relation (\ref{trade-off3}) as  the uncertainty relation between the measurement errors.  In this subsection, we  show that it can be interpreted as the uncertainty relation between the measurement error and   back-action of the measurement~\cite{Heisenberg,Fuchs,Banaszek-Devetak,Ozawa-2,Busch-Heinomen-Lahti}.
Let us suppose that $\hat{\rho}'$ is a state immediately after the measurement of $\hat A = \bm n_A \cdot \hat{\bm \sigma}$ for   the premeasurement state $\hat \rho$. If the measurement of $\hat A$ is described by measurement operators $\{ \hat M_k \}$, we can write $\hat \rho'$ as $\hat \rho' = \sum_k \hat M_k \hat \rho \hat M_k'^{\dagger}$.  For simplicity, we assume that the number of measurement outcomes is $2$: $k=1,2$.

To identify the disturbance of $\hat B = \bm n_B \cdot \hat{\bm \sigma}$ caused by the measurement of $\hat A$, we consider how much information about $\hat B$ for the premeasurement state $\hat \rho$ remains in post-measurement state $\hat{\rho}'$.
We characterize this by considering how much information on $\hat{\rho}$  can be obtained by performing the projection measurement of $\hat B$ for $\hat{\rho}'$.  Note that we can regard the projection measurement of $\hat B$ on $\hat \rho'$ described by the POVM  as the measurement of $\hat \rho$ described by the POVM $\{ \sum_k \hat M_k^{\dagger} \hat P_B(+) \hat M_k, \ \sum_k \hat M_k^{\dagger} \hat P_B(-) \hat M_k \}$, where $\bm n_B \cdot \hat{\bm \sigma} = \hat P_B(+) - \hat P_B(-)$. The joint operation of measurement $\hat A$   followed by measurement $\hat B$ can be described by a  POVM $\{ \hat{E}(i,j) \}$, where
\begin{equation}
\hat{E}(i,j) \equiv \hat M_i^{\dagger} \hat P_B(j) \hat M_i.
\end{equation}
We can construct the marginal POVMs as
\begin{equation}
\sum_j \hat E(i,j) = \hat M_i^{\dagger} \hat M_i, \ \sum_i \hat E(i,j) = \sum_i \hat M_k^{\dagger} \hat P_B (j) \hat M_k. 
\end{equation}

It is possible to interpret $1- \chi_B$ as a measure of  the back-action of $\hat B$ caused by measurement of $\hat A$.
Defining the measurement error of $\hat A$ as $\varepsilon_A \equiv (1/\chi_A) - 1$ and the back-action of the measurement on $\hat B$ as $d_B \equiv (1/\chi_B) - 1$, we can obtain the  trade-off relation between the error and  back-action based on inequality (\ref{trade-off3}).

\textbf{Theorem 11 }
(\textit{uncertainty relation between measurement error and   back-action}).
\begin{equation}
\varepsilon_A d_B \geq \sin^2 \theta
\label{e-b-trade-off}
\end{equation}

We  note that a non-selective measurement process for $\hat A$ can simulate the decoherence caused by the environment. 
In this case, the trade-off relation (\ref{e-b-trade-off}) gives a lower bound on the back-action of $\hat B$ in the presence of  decoherence characterized by $\chi_A$.

%%%%%%%%%%%%%%%%%%%%%%%%%%%%%%%%%%%%%%%%%%%%%%%%%%%%%%%%%%%%%%%%%%%%%%%%%%%%%%%%%%%%%%%%%%%%%%%%%%%%%%%%%%%%%%%%%%%%

\subsection{No-cloning Inequality}

Another application of the trade-off relation is the derivation of a no-cloning inequality. We consider a quantum cloning process from qubit system $\rm P$ to qubit system $\rm Q$ described as follows:  Let $\hat \rho$ be an  unknown density operator of system $\rm P$ to be cloned,  $\hat \rho_0$ be that of system $\rm Q$ as a blank reference state, and $\hat \rho_{\rm env}$ be that of the environment.  The density operator of the total system is initially given by $\hat \rho \otimes \hat \rho_0 \otimes \hat \rho_{\rm env}$, and  becomes $\hat U \hat \rho \otimes \hat \rho_0 \otimes \hat \rho_{\rm env} \hat U^{\dagger}$ after unitary evolution $\hat U$.  We define  $\hat \rho_{\rm P} \equiv {\rm tr_{Q,env}} (\hat U \hat \rho \otimes \hat \rho_0 \otimes \rho_{\rm env} \hat U^{\dagger})$ and $\hat \rho_{\rm Q} \equiv {\rm tr_{P,env}} ( \hat U \hat \rho \otimes \hat \rho_0 \otimes \hat \rho_{\rm env} \hat U^{\dagger} )$.  We can write $\hat \rho_{\rm P}$ and $\hat \rho_{\rm Q}$ in the operator-sum representation as $\hat \rho_{\rm P} = \sum_k \hat M_k \hat \rho \hat M_k^{\dagger}$ and $\hat \rho_{\rm Q} = \sum_k \hat M'_k \hat \rho \hat M_k'^{\dagger}$.

The no-cloning theorem~\cite{Wootters-Zurek,Dieks,Barnum} states that there exists no unitary operator $\hat U$ that satisfies $\hat \rho_{\rm P} = \hat \rho_{\rm Q} = \hat \rho$ for arbitrary input state $\hat \rho$. If $\hat U$ is the identity operator, then all information about $\hat \rho$ remains in system $\rm P$, and no information is transferred into system $\rm Q$; $\hat \rho_{\rm P} = \hat \rho$ and $\hat \rho_{\rm Q} = \hat \rho_0$.  As another special case, if $\hat U$ describes  the swapping operation between $\rm P$ and $\rm Q$ (i.e., $\hat U \hat \rho \otimes \hat \rho_0 \otimes \hat \rho_{\rm env} \hat U^{\dagger} = \hat \rho_0 \otimes \hat \rho \otimes \hat \rho_{\rm env}$), then all information about $\hat \rho$ is transferred into $\rm Q$ with no information left in $\rm P$. Intermediate cases between the identity operation and the swapping operation can be quantitatively analyzed by the no-cloning inequality~\cite{Cerf}.

We derive here another simple no-cloning inequality based on the trade-off relation.  We first consider how much information about $\hat \rho$ remains in $\hat \rho_{\rm P}$.  We can characterize this by considering how much information about $\bm n \cdot \hat{\bm \sigma}$ of $\hat \rho$ can be obtained by the measurement of $\bm n \cdot \hat{\bm \sigma}$ on $\hat \rho_{\rm P}$.  We can regard the measurement of $\bm n \cdot \hat{\bm \sigma}$ on $\hat \rho_{\rm P}$ as the measurement described by the POVM $\textbf{E}_{\rm P} (\bm n) = \{ \sum_k \hat M_k^{\dagger} \hat P(+; \bm n) \hat M_k, \ \sum_k \hat M_k^{\dagger} \hat P(-; \bm n) \hat M_k \}$ on $\hat \rho$, where $\bm n \cdot \hat{\bm \sigma} = P(+; \bm n) - P(-; \bm n)$.  We can thus characterize the amount of information that remains in $\rm P$ by the accuracy parameter $\chi (\bm n; \textbf{E}_{\rm P}(\bm n))$.
Similarly,  we can consider how much information about $\hat \rho$ is transferred into $\hat \rho_{\rm Q}$.  We characterize this by considering how much information about $\bm n \cdot \hat{\bm \sigma}$ of $\hat \rho$ can be obtained by the measurement of $\bm n \cdot \hat{\bm \sigma}$ on $\hat \rho_{\rm Q}$.  We can regard the measurement of $\bm n \cdot \hat{\bm \sigma}$ on $\hat \rho_{\rm Q}$ as the measurement described by the POVM $\textbf{E}_{\rm Q} (\bm n) = \{ \sum_k \hat M_k'^{\dagger} \hat P(+; \bm n) \hat M'_k, \ \sum_k \hat M_k'^{\dagger} \hat P(-; \bm n) \hat M'_k \}$ on $\hat \rho$.  We thus characterize the amount of information which is transferred from $\rm P$ to $\rm Q$ by the accuracy parameter $\chi (\bm n; \textbf{E}_{\rm Q}(\bm n))$.  For mathematical convenience, we use $\varepsilon_{\rm P} (\bm n) \equiv \varepsilon (\bm n;\textbf{E}_{\rm P}(\bm n))$ and $\varepsilon_{\rm Q} (\bm n) \equiv \varepsilon (\bm n; \textbf{E}_{\rm Q}(\bm n))$, instead of $\chi (\bm n;\textbf{E}_{\rm P}(\bm n))$ and $\chi (\bm n;\textbf{E}_{\rm Q}(\bm n))$, to derive our no-cloning inequality.  The amount of information about $\hat \rho$ which remains in $\rm P$ is characterized by $\varepsilon_{\rm P} (\bm n)$ averaged over all directions, and the amount of information about $\hat \rho$ which is transferred into $\rm Q$ is characterized by $\varepsilon_{\rm Q} (\bm n)$ averaged over all directions.

\textbf{Definition 4 }
(\textit{cloning parameter}).
We define the cloning parameters $C_{\rm P}$ and $C_{\rm Q}$ as 
\begin{equation}
C_{\rm P} \equiv \int_{| \bm n | =1} \varepsilon_{\rm P} (\bm n) \frac{d^3 \bm n}{4 \pi} , \ C_{\rm Q} \equiv \int_{| \bm n| =1}  \varepsilon_{\rm Q} (\bm n) \frac{d^3 \bm n}{4 \pi}. 
\end{equation}

Since $0 \leq \varepsilon_{\rm P} (\bm n) \leq \infty$ and $0 \leq \varepsilon_{\rm Q} (\bm n) \leq \infty$, the cloning parameters satisfy
\begin{equation}
0 \leq C_{\rm P} \leq \infty, \ 0 \leq C_{\rm Q} \leq \infty.
\end{equation}
The cloning parameters depend only on $\hat \rho_0$, $\hat \rho_{\rm env}$, and  $\hat U$, and characterize the performance of the cloning machine $\{ \hat U, \hat \rho_0, \hat \rho_{\rm env} \}$.  The smaller $C_{\rm P}$ is, the more information about $\hat \rho$ remains in system $\rm P$, while the smaller $C_{\rm Q}$ is, the more information about $\hat \rho$ is transferred into system $\rm Q$ by the cloning machine.  For example, if $\hat U$ is the identity operator, then  $C_{\rm P}=0$ and $C_{\rm Q}= \infty$ hold, which implies that all information about $\hat \rho$ is left in system $\rm P$.  On the other hand, if $\hat U$ describes  the swapping operation between $\rm P$ and $\rm Q$, then $C_{\rm P}= \infty$ and $C_{\rm Q} = 0$ hold.  For intermediate cases between them, the following no-cloning inequality between $C_{\rm P}$ and $C_{\rm Q}$ can be derived from trade-off relation~(\ref{trade-off3}).

\textbf{Theorem 12 }
(\textit{no-cloning inequality}):
\begin{equation}
C_{\rm P} C_{\rm Q} \geq \frac{2}{3}.
\label{no-cloning}
\end{equation}

\textbf{Proof. }
It can be shown that there exists a POVM $\textbf{E}(\bm n, \bm n') = \{ \hat E (i,j;\bm n, \bm n') \}$, with $i,j = \pm$,  satisfying
\begin{equation}
\begin{split}
{\rm tr} \left( (\hat P (i; \bm n) \otimes \hat I \otimes \hat I) (\hat I \otimes  \hat P (j; \bm n')\otimes \hat I ) \hat U \hat \rho \otimes \hat \rho_0 \otimes \hat \rho_{\rm env} \hat U^{\dagger} \right) \\
= {\rm tr} \left( \hat E(i,j; \bm n, \bm n') \hat \rho \right).
\end{split}
\end{equation}
We can also show that $\textbf{E}_{\rm P}(\bm n)$ and $\textbf{E}_{\rm Q}(\bm n')$ are its marginal POVMs.
From inequality~(\ref{data-processing5}) and the trade-off relation~(\ref{trade-off3}), we obtain
\begin{equation}
\begin{split}
\varepsilon_{\rm P} (\bm n) \varepsilon_{\rm Q} (\bm n') &\geq  \varepsilon ( \bm n; \textbf{E}(\bm n, \bm n')) \varepsilon (\bm n' ; \textbf{E}(\bm n, \bm n'))  \\
&\geq 1 - ( \bm n \cdot \bm n')^2.
\end{split}
\label{c}
\end{equation}
Averaging (\ref{c}) over   all directions and using
\begin{equation}
\int_{|\bm n | =1}  \int_{| \bm n' | =1} \varepsilon_{\rm P} (\bm n) \varepsilon_{\rm Q} (\bm n') \frac{d^3 \bm n}{4 \pi} \frac{d^3 \bm n'}{4 \pi}= C_{\rm P} C_{\rm Q},
\end{equation}
\begin{equation}
\int_{|\bm n | =1} \int_{| \bm n' | =1}  (1 - ( \bm n \cdot \bm n')^2 ) \frac{d^3 \bm n}{4 \pi}\frac{d^3 \bm n'}{4 \pi} = \frac{2}{3},
\end{equation}
we obtain (\ref{no-cloning}).---

Inequality (\ref{no-cloning}) represents the trade-off relation between the information remaining in the original system $\rm P$ and the information transferred to the reference system $\rm Q$.  The impossibility of achieving $C_{\rm P} = C_{\rm Q} = 0$ implies the no-cloning theorem.  Note that if $C_{\rm Q} \to 0$, then $C_{\rm P} \to \infty$, which implies that if a cloning machine transfers all of the information about $\hat \rho$ into system $\rm Q$, then no information can be left in system $\rm P$.

%%%%%%%%%%%%%%%%%%%%%%%%%%%%%%%%%%%%%%%%%%%%%%%%%%%%%%%%%%%%%%%%%%%%%%%%%%%%%%%%%%%%%%%%%%%%%%%%%%%%%%%%%%%%%

\subsection{Quantum State Tomography}

We next apply our framework to  quantum-state tomography~\cite{Banaszek,Hradil,Rehacek-2,Thew,James}.  As shown in Sec. V\hspace{-.1em}I,  characterization of the measurement accuracy by the accuracy matrix is closely related to the asymptotic accuracy of the maximum-likelihood estimation which is considered to be the standard scheme for quantum-state tomography.

We first consider the standard strategy to estimate the three components of  Bloch vector $\bm s_0$.
We divide $N$ identically prepared samples into three groups in the ratio $1 : 1 : 1$, and   measure $\hat \sigma_x$ for the first group,   $\hat \sigma_y$ for the second group, and  $\hat \sigma_z$ for the third group.
As $N$ increases, this scheme becomes asymptotically described by  POVM consisting of six operators:
\begin{equation}
\begin{split}
\hat E_1 = \frac{1}{6} (\hat I + \hat \sigma_x), \ \hat E_2 = \frac{1}{6} (\hat I - \hat \sigma_x), \\
\hat E_3 = \frac{1}{6} (\hat I + \hat \sigma_y), \ \hat E_4 = \frac{1}{6} (\hat I - \hat \sigma_y), \\
\hat E_5 = \frac{1}{6} (\hat I + \hat \sigma_z), \ \hat E_6 = \frac{1}{6} (\hat I - \hat \sigma_z).
\end{split}
\label{tomography1}
\end{equation}
We can reconstruct the quantum state by  quantum-state tomography  and hence reconstruct the probability distributions in all directions.  In fact, the accuracy matrix for the standard tomography (\ref{tomography1}) is given by 
\begin{equation}
\chi (\textbf{E})=
\left( 
\begin{array}{ccc}
1/3 & 0 & 0 \\
0 & 1/3 & 0 \\
0 & 0 & 1/3 
\end{array} 
\right) = \frac{1}{3} I_3,
\label{tomography-accuracy}
\end{equation}
which attains the upper bound of the inequality ${\rm Sp} (\chi(\textbf{E})) \leq 1$.
This expression manifestly shows that the reconstructive subspace of the standard quantum state tomography is $\mathbb{R}^3$ and that the accuracy of the tomography is optimal and symmetric in the sense discussed in Sec.~I\hspace{-.1em}I\hspace{-.1em}I B.

We next consider the minimal qubit tomography.  \v{R}eh\'{a}\v{c}ek \textit{et al.} have shown that the following four measured probabilities are just enough to estimate the Bloch vector~\cite{Rehacek-2}:
\begin{equation}
\hat E_k = \frac{1}{4} (1 + \bm a_k \cdot \hat{\bm \sigma}) \ (k=1,2,3,4),
\label{tomography2}
\end{equation} 
where
\begin{equation}
\begin{split}
\bm a_1 = \frac{1}{\sqrt{3}}
\left( 
\begin{array}{c}
1 \\
1 \\
1
\end{array} 
\right), \ 
\bm a_2 = \frac{1}{\sqrt{3}}
\left( 
\begin{array}{c}
1 \\
-1 \\
-1
\end{array} 
\right), \\
\bm a_3 = \frac{1}{\sqrt{3}}
\left( 
\begin{array}{c}
-1 \\
1 \\
-1
\end{array} 
\right), \ 
\bm a_4 = \frac{1}{\sqrt{3}}
\left( 
\begin{array}{c}
-1 \\
-1 \\
1
\end{array} 
\right).
\end{split}
\end{equation}
The minimal qubit tomography  is also optimal and symmetric, in the sense that the corresponding accuracy matrix is 
again given by (\ref{tomography-accuracy}).  Note that the POVM $\textbf{E}$ satisfying $V(\textbf{E})=\mathbb{R}^3$ can be regarded as tomographically complete~\cite{James}.

%%%%%%%%%%%%%%%%%%%%%%%%%%%%%%%%%%%%%%%%%%%%%%%%%%%%%%%%%%%%%%%%%%%%%%%%%%%%%%%%%%%%%%%%%%%%%%%%%%%%%%%%%%%%%%%%%%%

\section{Maximum-likelihood Estimation and the Fisher Information}

In this section, we point out a close connection between the accuracy matrix and the Fisher information~\cite{Fisher,Lehmann}.
We consider the quantum measurements described by the POVM $\textbf{E} = \{ \hat E_k \}$ for each of $N$ ($< \infty$) samples prepared in the same unknown state $\hat \rho$.  Note that $\hat E_k = r_k (\hat I + \bm v_k \cdot \hat{\bm \sigma})$.  Our task is to estimate the Bloch vector $\bm s_0$ by   maximum-likelihood estimation.  
We denote  $\bm s^{\ast}$ as the maximum-likelihood estimator of $\bm s_0$ from $N$ measurement outcomes.

The asymptotic accuracy of  maximum-likelihood estimation is characterized by the Fisher information.  In our situation, the Fisher information takes the matrix form given by
\begin{equation}
I_{ij} \equiv - \sum_k q_k \frac{\partial^2 \ln f_k (\bm s) }{\partial (\bm s)_i \partial (\bm s)_j} \biggr|_{\bm s = \bm s_0} = \sum_k \frac{r_k^2}{q_k}(\bm v_k)_i (\bm v_k)_j,
\label{Fisher1}
\end{equation}
or equivalently,
\begin{equation}
I = \sum_k \frac{r_k^2}{q_k} \bm v_k \bm v_k^{\rm T}.
\label{Fisher2}
\end{equation}
Note that $I$ is a $3 \times 3$ positive and Hermitian matrix, and that the support of $I$ coincides with that of $\chi(\textbf E)$.

Focusing on a  particular direction $\bm n$, we can reduce the Fisher information content to
\begin{equation}
I(\bm n) \equiv \frac{1}{\bm n \cdot I^{-1} \bm n}.
\label{Fisher3}
\end{equation}

The greater the Fisher information, the more information we can extract from the measurement outcome.  In the case of $I(\bm n) = 0$, the variance of the estimator $\bm n \cdot \bm s^{\ast}$ diverges, so we cannot gain any information about the probability distribution in direction $\bm n$.  This is the case of $\bm n$  not being in any reconstructive direction.

Replacing $q_k$ by $r_k$ in the Fisher information (\ref{Fisher1}) or (\ref{Fisher2}), we can obtain  the accuracy matrix in Eq.~(\ref{AM1}) or (\ref{AM2}). Note that  $r_k$ is the average of $q_k$ over the entire Bloch sphere.  The trade-off relations (\ref{trade-off3}), (\ref{trade-off4}), and (\ref{trade-off5}) can thus be interpreted as the trade-off relations  between the asymptotic accuracy of the maximum-likelihood estimation of the probability distributions of observables.  A finite number of samples only gives us  imperfect information about the probability distribution of an observable for an unknown state. As we have shown~\cite{Kurotani}, this imperfection  further deteriorates in the case of simultaneous estimation due to the noncommutability of the observables. 

Figure 2 shows the results of simulations for the value of the maximum-likelihood estimators $p(+;\bm n_A)^\ast$ (red curves) and $p(+; \bm n_B)^\ast$ (blue curves) in the the case of an optimal nonideal joint POVM which satisfies the equality in (\ref{trade-off3}) or (\ref{trade-off4}) with
\begin{equation}
\chi_A = 1/10 = 0.10, \ \chi_B = 36/37  \simeq 0.97
\end{equation}
and
\begin{equation}
\bm n_A = \left(
\begin{array}{c}
0 \\
0 \\
1
\end{array}
\right), \
\bm n_B = \left(
\begin{array}{c}
1/2 \\
0 \\
\sqrt{3}/2
\end{array}
\right), \ \hat \rho = \frac{\hat I + \hat \sigma_x}{2}.
\end{equation}

\begin{figure}[htbp]
\begin{center}
\includegraphics[width=70mm]{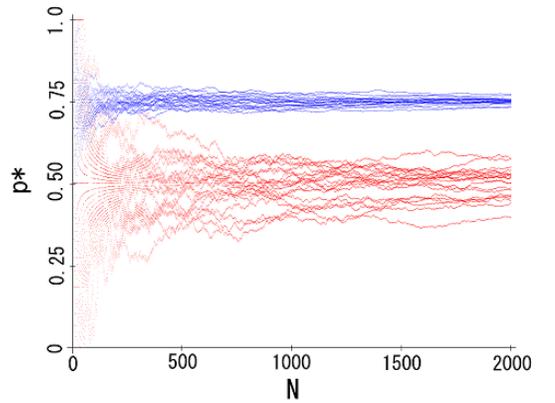}
\end{center}
\caption{(Color online) Maximum-likelihood estimators $p(+;\bm n_A)^\ast$ (red curves) and $p(+; \bm n_B)^\ast$ (blue curves) for the case of $\theta=\pi/6$, $\chi_A = 0.10$, and $\chi_B\simeq 0.97$. The abscissa indicates sample number $N$ and the ordinate indicates the value of the  estimators.  The number of simulations is $20$.} 
\label{fig:simulation}
\end{figure}

Let us next consider a simple estimation  scheme by dividing $N$ prepared samples into two groups in the ratio $\xi : 1-\xi \ (0<\xi<1)$ and performing a nonideal measurement of  $\bm n_A \cdot \hat{\bm \sigma}$ by the POVM $\{ \hat E (\pm; \bm n_A)  \}$ with accuracy $\chi_A$ for the former group, and similarly we perform a nonideal measurement of $\bm n_B \cdot \hat{\bm \sigma}$ by the POVM $\{ \hat E (\pm; \bm n_B) \}$ with accuracy $\chi_B$ for the latter group (see also example 3 in Sec.~I\hspace{-.1em}I).  This measurement can asymptotically be described by the POVM
\begin{equation}
\textbf{E} = \{ \xi \hat E (\pm \bm n_A), (1-\xi ) \hat E (\pm \bm n_B) \},
\end{equation}
whose accuracy matrix is
\begin{equation}
\chi(\textbf{E}) = \xi \chi_A \bm n_A \bm n_A^{\rm T} + (1-\xi) \chi_B \bm n_B \bm n_B^{\rm T}.
\end{equation}
From Eq.(\ref{accuracy2}) in Sec.~I\hspace{-.1em}I,  the  accuracy parameters in  directions $\bm n_A$ and $\bm n_B$ are given by 
\begin{equation}
\chi (\bm n_A; \textbf{E}) = \xi \chi_A, \ \chi (\bm n_B; \textbf{E}) = (1- \xi) \chi_B,
\end{equation}
and thus
\begin{equation}
\chi (\bm n_A; \textbf{E}) + \chi (\bm n_B; \textbf{E}) \leq 1.
\end{equation} 
We can therefore conclude that a simultaneous measurement has the advantage over this simple method in that the former can access the domain $\chi_A + \chi_B > 1$ for $\theta \ne \pi/2$, i.e., domain Q in Fig.1.

%%%%%%%%%%%%%%%%%%%%%%%%%%%%%%%%%%%%%%%%%%%%%%%%%%%%%%%%%%%%%%%%%%%%%%%%%%%%%%%%%%%%%%%%%%%%%%%%%%%%%%%%%%%%%%

\section{Summary and Discussion}

Projection measurements cannot always be implemented  experimentally.  This raises the question of how  accurately  we can obtain  information about observables from a given imperfect measurement scheme.  To quantitatively characterize such measurement accuracy, we have introduced  the $3 \times 3$ accuracy matrix $\chi (\textbf{E})$, with $\textbf{E} = \{ \hat E_k \}$ being the corresponding POVM.   

We have considered the accuracy matrix of the most general class of measurements of a qubit system:  generalized simultaneous measurements  including nonideal joint measurements and quantum-state tomography.  From the outcomes of   generalized simultaneous measurements, we can obtain  information about more than one observable.

In terms of the accuracy matrix, we have defined accuracy parameter $\chi (\bm n; \textbf{E})$ and error parameter $\varepsilon (\bm n; \textbf{E})$ for a direction of  $\bm n$ corresponding to the observable $\bm n \cdot \hat{\bm \sigma}$. These parameters satisfy   $0 \leq \chi(\bm n ; \textbf{E}) \leq 1$ and $0 \leq \varepsilon (\bm n; \textbf{E}) \leq \infty$. If $\chi(\bm n ; \textbf{E}) = 1$, or equivalently $\varepsilon (\bm n; \textbf{E})=0$, the measurement  is equivalent to the projection measurement of $\bm n \cdot \hat{\bm \sigma}$.  On the other hand, if $\chi(\bm n ; \textbf{E}) = 0$, or equivalently $\varepsilon (\bm n; \textbf{E})=\infty$, we cannot obtain any  information about the measured system by this measurement.

The accuracy matrix and accuracy parameters give us information about observables for which we can reconstruct the probability distribution from  the measured distribution $\{ q_k \}$, where $q_k \equiv {\rm tr}(\hat \rho \hat E_k) $. In fact, we can reconstruct the probability distribution of observable $\bm n \cdot \hat{\bm \sigma}$ if and only if $\chi(\bm n; \textbf{E}) \neq 0$, or equivalently $\varepsilon (\bm n; \textbf{E}) < \infty$.  In other words, the direction $\bm n$ is a reconstructive direction if and only if $\bm n \in V(\textbf{E})$, where the subspace $V(\textbf{E})$ of $\mathbb R^3$ is spanned by the eigenvectors of $\chi (\textbf{E})$ corresponding to nonzero eigenvalues.

The main results of this paper are  trade-off relations  (\ref{trade-off3}), (\ref{trade-off4}), and (\ref{trade-off5}) between the accuracy parameters and the error parameters.  We can interpret them  as the uncertainty relations between measurement errors in generalized simultaneous measurements; the more information we obtain about an observable, the less information we can access about the other noncommuting observable.  Trade-off relation~(\ref{trade-off3}) can also be interpreted as the uncertainty relation between the measurement error and   back-action of measurement as formulated in inequality~(\ref{e-b-trade-off}).

The new  no-cloning inequality in~(\ref{no-cloning}) is derived from the trade-off relations.  To derive this, we have introduced the cloning parameters $C_{\rm P}$ and $C_{\rm Q}$, where $\rm P$ indicates the system to be cloned  and  $\rm Q$ indicates the  blank reference system. Let $\hat \rho$ be the pre-cloned state of system $\rm P$. After a cloning operation, all the information about $\hat \rho$ remains in system $\rm P$  if and only if  $C_{\rm P}=0$, and the information about $\hat \rho$ is completely transferred to system $\rm Q$ if and only if $C_{\rm Q}=0$.  The impossibility of attaining  $C_{\rm P} = C_{\rm Q} = 0$ implies the no-cloning theorem.  The condition of the equality  in our no-cloning inequality~(\ref{no-cloning}) has yet to be understood.

We have also applied the trade-off relations to  analyze the efficiency of quantum-state tomography.  The accuracy matrix of the standard qubit-state tomography or the minimal qubit tomography is given by $\chi(\textbf{E})= I_3/3$ with $I_3$ being the $3 \times 3$ identity matrix, which implies that the efficiency of quantum-state tomography is optimal and symmetric.

We have pointed out a close relationship between the accuracy matrix and the Fisher information.  We have also shown that the trade-off relations can be interpreted as being those concerning the accuracy of the maximum-likelihood estimators of the probability distributions of noncommuting observables.  

While we focus on the spin-1/2 system in the present paper, many results can be generalized for higher-dimensional systems. We conclude this paper by outlining such generalization.

In the case of a $d$-dimensional system ($d \geq 3$), the parametrization of the Hermitian operator $\hat E$ is given by
\begin{equation}
\hat E = r (\hat I + \sqrt{d-1} \bm v \cdot \hat{\bm \lambda} ),
\label{d-parameterize}
\end{equation}
where $r$ is a real  number, $\bm v$ is a $d^2-1$-dimensional real vector, and $\hat{\bm \lambda} = (\hat \lambda_1, \hat \lambda_2, \cdots, \hat \lambda_{d^2-1})$ is the elements of the Lie algebra of SU($d$) satisfying ${\rm tr}(\hat \lambda_i)=0$ and ${\rm tr} (\hat \lambda_i \hat \lambda_j) = d\delta_{ij}$ with $\delta_{ij}$ being the Kronecker delta.  The necessary and sufficient condition for $\hat E$ to be a positive operator is given by  $r > 0$ and $S_m(\bm v) \geq 0$ ($m=2,\cdots, d$), where $S_m (\bm v)$ is an  $m$th-degree polynomial for $\bm v$~~\cite{Kimura,Byrd}.  The condition for $m=2$ is given by $S_2 (\bm v) \equiv d(d-1) r^2 (1-| \bm v |^2 )/2 \geq 0$, which is equivalent to 
\begin{equation}
| \bm v | \leq 1.
\label{positivity2}
\end{equation}  For $m = 3$, $S_3 (\bm v)$ is given by
\begin{equation}
\begin{split}
S_3 (\bm v) &\equiv \frac{1}{6} d(d-1)(d-2)r^3 \\ 
&\times \left( 1 - 3 |\bm v|^2 +\frac{\sqrt{d-1}}{d-2} \sum_{ijk} d_{ijk} (\bm v)_i ( \bm v)_j ( \bm v)_k \right), 
\end{split}
\end{equation}
where $d_{ijk}$ is defined as $\{ \hat \lambda_i , \hat \lambda_j \} = 2\delta_{ij}\hat I + \sum_k d_{ijk}\hat \lambda_k$ with $\{ \hat A, \hat B \} \equiv \hat A \hat B + \hat B \hat A$.  We note that if $\hat E$ is a rank-$1$ projection operator, then $| \bm v | =1$.  However, the Hermitian operator $\hat E$ with $|\bm v |=1$ is not necessarily a positive operator.

The accuracy matrix for a $d$-dimensional system assumes the same form as Eq.~(\ref{AM2}) using parametrization~(\ref{d-parameterize}).  In this case, $\chi (\textbf{E})$ is a $d^2-1$ square matrix.
Moreover, we can define the accuracy parameter and the error parameter according to Eqs.~(\ref{A-parameter}) and~(\ref{E-parameter}), respectively.  Using condition~(\ref{positivity2}), we can derive trade-off relations (\ref{trade-off3}) and (\ref{trade-off4}) for a $d$-dimensional system.  In this sense, the trade-off relations serve as  universal uncertainty relations holding true for all finite-dimensional systems.  

However, bounds of trade-off relations~(\ref{trade-off3}) and (\ref{trade-off4}) would not necessarily be able to be reached  for $d \geq 3$, because $r>0$ and $|\bm v | =1$ are not sufficient for positivity of the POVM. Moreover, while the accuracy parameter $\chi (\bm n; \textbf{E})$ for $d=2$ characterizes the measurement accuracy of spin observables $\bm n \cdot \hat{\bm \sigma}$,  the accuracy parameter $\chi (\bm n; \textbf{E})$ for $d \geq 3$   cannot characterize the measurement accuracy of, for example,  the spin-$d$ observable $\hat J_z$; it only characterizes the accuracy of a rank-$1$ projection operator.  Therefore the results of this paper based on $\chi (\bm n; \textbf{E})$ cannot be applied straightforwardly  for $d \geq 3$.   A full investigation of this problem is underway.

%%%%%%%%%%%%%%%%%%%%%%%%%%%%%%%%%%%%%%%%%%%%%%%%%%%%%%%%%%%%%%%%%%%%%%%%%%%%%%%%%%%%%%%%%%%%%%%%%%%%%%%%%%%%%%%%%%

\appendix

\section{Proof of Theorem 10}

We prove the case of $\alpha = A$.  For simplicity of notation, we define that $\hat E_1 \equiv \hat E (+,+)$, $\hat E_2 \equiv \hat E(+,-)$, $\hat E_3 \equiv \hat E(-,+)$, and $\hat E_4 \equiv \hat E(-,-)$.  The accuracy matrix is given by
\begin{equation}
\chi(\textbf{E}) = \sum_{k=1}^4 r_k | \bm v_k |^2,
\end{equation}
and the accuracy parameter in direction $\bm n_A$ is
\begin{equation}
\chi_A = \left( \bm n_A \cdot (\chi(\textbf{E})^{-1}) \bm n_A \right)^{-1}.
\end{equation}
The marginal POVM $\textbf{E}_A$ is 
\begin{equation}
\hat E_A(+)  = (r_1 + r_2) \left( \hat I + \frac{r_1 \bm v_1 + r_2 \bm v_2}{r_1 + r_2} \cdot \hat{\bm \sigma} \right),
\end{equation}
\begin{equation}
\hat E_A(-) = (r_3 + r_4) \left( \hat I + \frac{r_3 \bm v_3 + r_4 \bm v_4}{r_3 + r_4} \cdot \hat{\bm \sigma} \right),
\end{equation}
and the marginal accuracy matrix is
\begin{equation}
\chi(\textbf{E}_A) = \chi'_A \bm n_A \bm n_A^{\rm T},
\end{equation}
where 
\begin{equation}
\chi'_A = \frac{|r_1 \bm v_1 + r_2 \bm v_2|^2}{r_1 + r_2} + \frac{|r_3 \bm v_3 + r_4 \bm v_4|^2}{r_3 + r_4}.
\end{equation}
Our objective is to show that $\chi_A = \chi'_A$.
For simplicity, we introduce the notation
\begin{equation}
\bm a_k \equiv \sqrt{r_k}\bm v_k \equiv 
\left( 
\begin{array}{c}
x_k \\
y_k \\
z_k
\end{array} 
\right).
\end{equation}
We can then write $\chi(\textbf{E})$ as
\begin{equation}
\chi (\textbf{E}) = \sum_{k=1}^4 \bm a_k  \bm a_k^{\rm T} \equiv \left( \bm \chi_x, \bm \chi_y, \bm \chi_z \right),
\end{equation}
where
\begin{equation}
\bm \chi_x = \sum_{k=1}^4 x_k \bm a_k, \ \bm \chi_y = \sum_{k=1}^4 y_k \bm a_k, \ \bm \chi_z = \sum_{k=1}^4 z_k \bm a_k.
\label{A1}
\end{equation}
Using Eq.~(\ref{A1}), we can calculate the determinant of $\chi(\textbf{E})$:
\begin{equation}
\begin{split}
\det \chi(\textbf{E}) &= \left( \sum_{k=1}^4 x_k \bm a_k \right) \cdot \left( \sum_{k=1}^4 y_k \bm a_k  \times \sum_{k=1}^4 z_k \bm a_k \right) \\
&= \sum_{k<l<m} \bigl[ \bm a_k \cdot (\bm a_l \times \bm a_m) \bigr]^2.
\end{split}
\end{equation}
On the other hand, the cofactor matrix of $\chi(\textbf{E})$ is
\begin{equation}
\tilde{\chi}(\textbf{E}) = 
\left( 
\begin{array}{ccc}
(\bm \chi_y \times \bm \chi_z)_x & (\bm \chi_y \times \bm \chi_z)_y  & (\bm \chi_y \times \bm \chi_z)_z \\
(\bm \chi_z \times \bm \chi_x)_x& (\bm \chi_z \times \bm \chi_x)_y & (\bm \chi_z \times \bm \chi_x)_z  \\
 (\bm \chi_x \times \bm \chi_y)_x & (\bm \chi_x \times \bm \chi_y)_y & (\bm \chi_x \times \bm \chi_y)_z
\end{array} 
\right).
\end{equation}
Therefore 
\begin{equation}
\tilde{\chi}(\textbf{E})_{ij} = \sum_{k<l} (\bm a_k \times \bm a_l)_i (\bm a_k \times \bm a_l)_j.
\end{equation}
The inverse matrix is given by
\begin{equation}
(\chi(\textbf{E})^{-1} )_{ij} = \frac{\sum_{k<l} (\bm a_k \times \bm a_l)_i (\bm a_k \times \bm a_l)_j}{\sum_{k<l<m} \bigl[ \bm a_k \cdot (\bm a_l \times \bm a_m) \bigr]^2}.
\end{equation}

Noting that
\begin{equation}
\begin{split}
&   (r_1 \bm v_1 + r_2 \bm v_2) \cdot \tilde{\chi}(\textbf{E}) (r_1 \bm v_1 + r_2 \bm v_2)  \\
&= \sum_{i,j} \sum_{k<l} (r_1 \bm v_1 + r_2 \bm v_2)_i (\bm a_k \times \bm a_l)_i (\bm a_k \times \bm a_l)_j (r_1 \bm v_1 + r_2 \bm v_2)_j \\
&= \sum_{k<l} \bigl[ (r_1 \bm v_1 + r_2 \bm v_2) \cdot (\bm a_i \times \bm a_j) \bigr]^2 \\
&= (r_1 + r_2) \left( \bigl[ \bm a_1 \cdot (\bm a_2 \times \bm a_3) \bigr]^2 + \bigl[ \bm a_1 \cdot (\bm a_2 \times \bm a_4) \bigr]^2 \right) 
\end{split}
\end{equation}
and
\begin{equation}
\bm n_A =\frac{r_1 \bm v_1 + r_2 \bm v_2}{|r_1 \bm v_1 + r_2 \bm v_2|},
\end{equation}
we obtain
\begin{equation}
\begin{split}
&  \bm n_A \cdot ( \chi(\textbf{E})^{-1} ) \bm n_A  \\
&= \frac{r_1 + r_2}{|r_1 \bm v_1 + r_2 \bm v_2|^2} \frac{\bigl[ \bm a_1 \cdot (\bm a_2 \times \bm a_3) \bigr]^2 + \bigl[ \bm a_1 \cdot (\bm a_2 \times \bm a_4) \bigr]^2}{ \det \chi (\textbf{E})}.
\end{split}
\end{equation}
Similarly, we can show that
\begin{equation}
\begin{split}
&  \bm n_A \cdot (\chi(\textbf{E})^{-1}) \bm n_A  \\
&= \frac{r_3 + r_4}{|r_3 \bm v_3 + r_4 \bm v_4|^2} \frac{\bigl[ \bm a_3 \cdot (\bm a_4 \times \bm a_1) \bigr]^2 + \bigl[ \bm a_3 \cdot (\bm a_4 \times \bm a_2) \bigr]^2}{ \det \chi (\textbf{E})}.
\end{split}
\end{equation}
Let us define
\begin{equation}
X \equiv \frac{\bigl[ \bm a_1 \cdot (\bm a_2 \times \bm a_3) \bigr]^2 + \bigl[ \bm a_1 \cdot (\bm a_2 \times \bm a_4) \bigr]^2}{ \det \chi (\textbf{E})},
\end{equation}
\begin{equation}
Y \equiv \frac{\bigl[ \bm a_3 \cdot (\bm a_4 \times \bm a_1) \bigr]^2 + \bigl[ \bm a_3 \cdot (\bm a_4 \times \bm a_2) \bigr]^2}{ \det \chi (\textbf{E}) },
\end{equation}
$t \equiv r_1 + r_2 = 1- ( r_3 + r_4)$, and $1/\eta \equiv |r_1 \bm v_1 + r_2 \bm v_2|^2 = |r_3 \bm v_3 + r_4 \bm v_4|^2$. Noting that $X+Y=1$ and
\begin{equation}
\bm n_A \cdot (\chi(\textbf{E})^{-1}) \bm n_A  = Xt\eta = Y(1-t) \eta,
\end{equation}
we obtain
\begin{equation}
X+t =1.
\end{equation}
We can thus conclude
\begin{equation}
\bm n_A \cdot (\chi(\textbf{E})^{-1}) \bm n_A  = t (1-t) \eta.
\end{equation}
Therefore
\begin{equation}
\begin{split}
\chi_A &= \frac{1}{\bm n_A \cdot (\chi(\textbf{E})^{-1}) \bm n_A } \\ 
&= \frac{1}{\eta (r_1 + r_2) (r_3 + r_4)} \\
&= \frac{1}{\eta} \left( \frac{1}{r_1 + r_2} + \frac{1}{r_3 + r_4} \right) \\
&= \frac{|r_1 \bm v_1 + r_2 \bm v_2|^2}{r_1 + r_2} + \frac{|r_3 \bm v_3 + r_4 \bm v_4|^2}{r_3 + r_4} \\
&= \chi'_A,
\end{split}
\end{equation}
which is our objective.

%%%%%%%%%%%%%%%%%%%%%%%%%%%%%%%%%%%%%%%%%%%%%%%%%%%%%%%%%%%%%%%%%%%%%%%%%%%%%%%%%%%%%%%%%%%%%%%%%%%%%%%%%%%%%%%%%

\begin{acknowledgments}
This work was supported by a Grant-in-Aid for Scientific Research (Grant No.\ 17071005) and by a 21st Century COE program at Tokyo Tech, ``Nanometer-Scale Quantum Physics'', from the Ministry of Education, Culture, Sports, Science and Technology of Japan. 

We thank Y. Watanabe for performing numerical  simulations in Sec. V\hspace{-.1em}I.
\end{acknowledgments}

%\newpage %Just because of unusual number of tables stacked at end
%\bibliography{References}% Produces the bibliography via BibTeX.

\end{document}